\pdfoutput=1 
\documentclass[
 reprint,
showpacs,
 amsmath,amssymb,
 aps,
 pre,
floatfix,
]{revtex4-1}
\usepackage{graphicx}
\usepackage{dcolumn}
\usepackage{bm}

\usepackage{epsfig} 
\DeclareGraphicsExtensions{.pdf,.jpeg,.png,.jpg,.eps,}
\usepackage{subfig}
\usepackage{multirow}
\usepackage{url}
\usepackage[format=plain, justification=centerlast]{caption}
\newcommand{\reffig}[1]{Fig.\ref{#1}}	
\newcommand{\reftab}[1]{Table \ref{#1}}	
\newcommand{\subreffig}[2]{Fig.\ref{#1}\subref{#2}}

\begin{document}

\preprint{APS/123-QED}

\title{Temporal Stable Community in Time-Varying Networks}

\author{Wenjing Wang}
 \email{wenjingwang16@fudan.edu.cn}
\author{Xiang Li}%
 \email{Corresponding author, lix@fudan.edu.cn}
\affiliation{%
 Adaptive Network and Control (CAN) Lab, Department of Electronic Engineering\\
Research Center of Smart Networks and Systems,\\
School of Information Science and Engineering, Fudan University, Shanghai, 200433, China
}%

\date{\today}

\begin{abstract}
Identifying community structure of a complex network provides insight to the interdependence between the network topology and emergent collective behaviors of networks, while detecting such invariant communities in a time-varying network is more challenging. In this paper, we define the temporal stable community and newly propose the concept of dynamic modularity to evaluate the stable community structures in time-varying networks, which is robust against small changes as verified by several empirical time-varying network datasets. Besides, using the volatility features of temporal stable communities in functional brain networks, we successfully differentiate the ADHD (Attention Deficit Hyperactivity Disorder) patients and healthy controls efficiently.
\end{abstract}
\maketitle


\section{\label{sec:level1}Introduction}

Network indices are an efficient tool to characterize the relationships among a huge number of entities in diverse fields such as sociology, biology and computer science\cite{newman2003structure,Chen2012Introduction}. Most real networks exhibit community structures, which represent groups of nodes and features having dense connectivity within the groups and relative sparser connectivity among the groups\cite{ISI:000176217700005,newman2006modularity}. For example, families or friends tend to form communities in online and contact-based social networks\cite{scott2017social}; researchers belonging to the same community in a collaborations network tend to have more frequent academic collaboration\cite{moody2004structure}; communities in functional brain networks are likely to group brain regions having similar functions\cite{sporns2016modular}. Uncovering communities in networks therefore is significant to understand the roles of such nodes in a community with a variety of applications\cite{fortunato2010community,malliaros2013clustering}.

A series of community detection methods have been developed for static networks, which mostly find an optimal solution according to the defined quality function. For example modularity maximization methods\cite{newman2016equivalence} and spectral clustering methods\cite{ISI:000249409100009} are recognized as classic algorithms of community detection in static networks. However, most real networks are dynamic in nature, $i.e.$, nodes and connectivity are in evolution over time\cite{holme2012temporal}. For example, new computers are added into the Internet continuously, and the brain networks can make adjustments in response to the external stimuli\cite{ISI:000322416000030}. Simply aggregating the information at different time points into a single network may overlook such essential features. Time-varying networks, where network structures at different time points are encoded by multiple layers, are drawing more and more attention\cite{pfitzner2013betweenness,zhang2014susceptible,zhang2015human,hou2016structural,liang2016identifying,li2017reconstruction}.

Recently several efforts have focused on the communities in time-varying networks. Ref.~\cite{aslak2018constrained} modeled the node-level inter-layer dependencies in temporal networks based on the neighborhood flow patterns within each layer, and identified intermittent communities by combining the Infomap method. Ref.~\cite{peixoto2015inferring} generalized the stochastic block models to characterize layered, edge-valued, and time-varying networks. The models were formulated in a nonparametric Bayesian framework which allows the identification of communities in real-world networks. Ref~.\cite{ghasemian2016detectability} derived a precise detectability threshold based on the dynamic stochastic block model, which depends on the rate of change and the strength of communities. Ref.~\cite{mucha2010community} developed a generalization of modularity to track the changes of communities in networks that evolve over time, have multiple types of links or have multiple scales. Refs.~\cite{zhang2012common,zhang2017finding} tried to find a common community structure across layers by computing a structure mathematically most similar to that in each single layer of the network. Although combining the information in different layers, these methods ignored the temporal patterns and temporal correlations of connectivity changes, and the detected community structure is not relevant to the time order of network layers.

In reality, the edges in networks are not independent of each other, and the temporal correlations of connectivity changes are related to the function specialization, information processing, social adhesion and systemic risk\cite{bassett2014cross}. Communities, as well as the edges inside and between communities in the same network, bear different temporal patterns. Ignoring such difference can mix nodes with different temporal dynamics together. In this paper, we quantitatively characterize the temporal changes of networks, and explicitly incorporate the temporal characteristics into the detection of stable communities in time-varying networks. The contributions of our work are threefold: (i) We develop the concept of relative volatility and the entropy of volatility which allow to quantify the degree to which the network changes over time. (ii) We define a concept of the stable community in time-varying networks and propose an efficient tool to evaluate the quality of the stable community structure, $i.e.$, the dynamic modularity. (iii) Based on the maximization of the dynamic modularity, we put forth the detection method of the stable community by extending the Louvain method. In addition, applications of the proposed method to the United States Congress voting data and the brain fMRI data suggest the effectiveness, significance and robustness of the stable community structure.


\section{\label{sec:leve2}Temporal Stable Community Detection in Time-Varying Networks}

The identification of stable communities can be considered as an optimization problem of dynamic modularity. In this section we first fix the notation and propose the concept of volatility, based on which we then define the dynamic modularity. In order to maximize the dynamic modularity, we extend the Louvain method to time-varying network settings. Meanwhile, procedures of significance test and robustness test for stable communities are put forward.

\subsection{\label{sec:leve21}Notation}
Consider a time-varying network $G_d=(G(t),t=1\cdots T)$. There are $N$ nodes and $T$ layers. In each layer, $G(t)=(V,E(t))$ stands for the network at time $t$, where the node set $V=\left\{v_1,v_2,\cdots, v_N\right\}$ remains unchanged over the time, and the edge set $E(t)=(e_{ij}(t), i,j=1,2,\cdots N, i\neq j)$ contains the edges active at time $t$. The weights of edges are recorded as the form of matrix $A(t)=[a_{ij}(t)]_{N\times N}$, where $a_{ij}(t)$ denotes the weight of edge $e_{ij}$ at time $t$. In this paper, we focus on a weighted undirected network, thus $A(t)$ is symmetric. The aim of the stable community detection is to divide $N$ nodes into $K$ disjoint communities by combining the network topology at different time and the temporal patterns of network changes. Moreover, the affiliation of each node to communities remains stable even though the connectivity among nodes changes over time. The vector $C=[C_1,C_2,\cdots,C_N]$ denotes the community set. If an edge links two nodes in the same community, it is referred to as an inner-community edge, otherwise an inter-community edge.

\subsection{\label{sec:leve22}Volatility Measures}
Essentially, the temporal changes of networks arise from the changes of edges including their appearance and disappearance, as well as the fluctuation of edge weights. We firstly introduce the concept of volatility and extend it to measure the extent of such variations in time-varying networks. The volatility of link $e_{ij}$ between node $i$ and node $j$, denoted as $V_{ij}$, is defined as\cite{figlewski1997forecasting}:
\begin{equation}\label{eq1}
\begin{split}
  V_{ij}= & \sqrt{T}std(\Delta_{ij}(t)) \\
  \Delta_{ij}(t)= & \frac{a_{ij}(t)-a_{ij}(t-1)}{a_{ij}(t-1)}, 1< t\leq T \\
\end{split}
\end{equation}
where $a_{ij}(t)$ represents the weight of edge $e_{ij}$ in network $G(t)$, and $std(\cdot)$ is the operator of standard deviation. The value of volatility $V_{ij}\in R$, and the larger the volatility is, the more severely the edge changes. In time-varying network $G_d$, the volatility of all edges are recorded in the volatility matrix $V_m=[V_{ij}]_{N\times N}$.

The degree to which an edge varies over time depends on not only itself, but also the degree to which the whole network changes. In order to compare the volatility of the same edge at different situations, we define the relative volatility, characterizing the variation extent of a single edge compared to the whole network. There are two types of relative volatility, $i.e.$, the pure relative volatility $V_{ratio}$ and the Pearson relative volatility $V_{pratio}$. The $V_{ratio}$ of edge $e_{ij}$ is defined as:
\begin{equation}\label{eq2}
\begin{split}
   V_{ratio}^{ij}= & \frac{V_{ij}}{V_g} \\
   \bm{g(t)}= & \frac{2}{N(N-1)}\sum_{i,j=1,i<j}^{i,j=N}\bm{a_{ij}(t)} \\
\end{split}
\end{equation}
where $\bm{g(t)}$ is the average weight series of all edges in the network $G_d$, and $V_g$ is the volatility of $\bm{g(t)}$.
The $V_{pratio}$ of edge $e_{ij}$ is defined as:
\begin{equation}\label{eq3}
  V_{pratio}^{ij}=V_{ratio}^{ij}\times corr(\bm{e_{ij}(t)},\bm{g(t)})
\end{equation}
where $corr(\bm{e_{ij}(t)},\bm{g(t)})$ denotes the Pearson correlation coefficient of the weight series of edge $e_{ij}$ and $\bm{g(t)}$.

Furthermore, we newly propose the volatility entropy, denoted as $H_v$, to describe the heterogeneity of edge weights. The entropy $H_v$ is defined as:
\begin{equation}\label{eq4}
  H_v=-\sum_{\hat{v}=min(\hat{V_{ij}})}^{\hat{v}=max(\hat{V_{ij}})}p(\hat{v})log(\hat{v})
\end{equation}
where $V_{ij}$ is the elements in the volatility matrix $V_m$. In order to calculate the entropy, the volatility $V_{ij}$ is rounded to integer $\hat{v}$, and $p(\hat{v})$ is the proportion of edges with the rounded volatility $\hat{v}$. The $H_v$ of all of edges in the whole network, inner-community edges, and inter-community edges are denoted as HV, HV\_inner and HV\_inter, respectively.
\subsection{\label{sec:leve23}Dynamic Modularity}
Extending the Newman-Girvan modularity in static networks\cite{Newman2004}, we define the dynamic modularity ($DQ$) for time-varying networks to describe the degree to which nodes form the stable communities during a course of time.
Based on the assumption that there exist stable community structures that are not affected by temporal changes of networks, temporal stable communities should satisfy the following conditions:

(i) One node belongs to one community alone and such relationship does not change over time.

(ii) Nodes in the same community are connected much more tightly than those belong to different communities. In a weighted network, the edge weight could be thought as a measure of the closeness of connectivity\cite{kumpula2007emergence}. Thus the average weight of inner-community edges should be larger than that of inter-community edges.

(iii) The weights of inter-community edges fluctuate much more severely than those of inner-community edges. Here we adopt the volatility introduced in section \ref{sec:leve2} to describe the extent of variation in edge weights. Consequently, the volatility of inter-community edges should be larger compared to inner-community edges.

Accordingly, the $DQ$ of a time-varying network $G_d=(G(t),t=1\cdots T)$ is defined as follows:
\begin{equation}\label{e5}
\begin{split}
   DQ= & \frac{\sum_{t}m_t[\frac{\sum_{i,j}(a_{ij}(t)-\gamma P_{ij})\delta(C_i,C_j)}{\sum_{i,j}{\delta(C_i,C_j)}}]}{\sum_{t}m_t} \\
     & -\frac{\sum_{i,j}V_{ij}\delta(C_i,C_j)}{\sum_{i,j}\delta(C_i,C_j)}
\end{split}
\end{equation}
where $m_t$ is the number of edges in network $G_t$, $\gamma$ is a structural resolution parameter, $P_{ij}(t)$ is the expected weight of the edge linking node $i$ and node $j$ in null models, i.e., randomized networks keeping some structural properties of the original networks unchanged, e.g., see the configuration model in Ref.~\cite{molloy1995critical, newman2001random}. If node $i$ and node $j$ are assigned to the same community, $\delta(C_i,C_j)$ is 1 otherwise 0. $\sum_{i,j}\delta(C_i,C_j)$ is the total number of inner-community edges.

The first part of the definition of $DQ$ in Eq.\eqref{e5} characterizes the closeness of connectivity inside communities. Different from the Newman-Girvan modularity, the average rather than the sum of weights is adopted in case that the size of communities is small, and then the sum of weights of inner-community edges could be smaller than that of inter-community edges.

The latter part of the definition in Eq.\eqref{e5} is the negative average of volatility of inner-community edges. The temporal properties of stable communities are explicitly defined through the difference of volatility  between inner-community edges and inter-community edges. With the volatility introduced into the definition of $DQ$, time-varying networks are treated as a whole rather than independent single layers.

\subsection{\label{sec:leve24}Modularity Maximization}
The partitions corresponding to the maximum of dynamic modularity are stable communities of the network during the period of time. We rewrite the definition of $DQ$ in Eq.\eqref{e5} as follows:
\begin{equation}\label{e6}
\begin{split}
Q_t=& \frac{\sum_{i,j}(a_{ij}(t)-\gamma\frac{s_i(t)s_j(t)}{2\omega(t)}-V_{ij})\delta(C_i,C_j)}{\sum_{i,j}\delta(C_i,C_j)}\\
= & \frac{\sum_{i,j}b_{ij}(t)\delta(C_i,C_j)}{\sum_{i,j}\delta(C_i,C_j)}\\
DQ=&\frac{\sum_{t}m_tQ_t}{\sum_{t}m_t}\\.
\end{split}
\end{equation}
where the configuration model\cite{molloy1995critical, newman2001random} is used as the null model, $s_i(t)$ and $s_j(t)$ denote the strength of node $i$ and node $j$, respectively,  $\omega(t)$ is the sum of edge weights in network $G(t)$. $b_{ij}(t)$ in network $G(t)$ is organized as the form of dynamic modularity matrix $DB(t)=[b_{ij}(t)]_{N\times N}$. The adjusted formula \eqref{e6} is similar to Newman-Girvan modularity in form. Therefore, we extend the Louvain method\cite{blondel2008fast} to time-varying network settings for the maximization of dynamic modularity, denoted as the extended Louvain method.

The Louvain method is a two-phase fast greedy optimization method for static networks. First, the method treats each node as a community and moves nodes into one of its adjacent communities to achieve the maximum increase in modularity. Second, it aggregates nodes in the same community and constructs a new network. These steps are repeated iteratively until a maximum modularity is attained. In a time-varying network, however, moving a node from its own community to another community could bring about the changes of $Q_t$ in each layer of the network. According to Eq.\eqref{e6}, the increase in $DQ$ of a time-varying network, denoted $dDQ$, is equal to the sum of increases in $Q_t$ of all layers : $dDQ=\frac{\sum_{t}m_tdQ_t}{\sum_{t}m_t}$, where $dQ_t$ denotes the increment of $Q_t$.

For computational convenience, the dynamic modularity contribution matrix, denoted $Hnm(t)$, is introduced: $Hnm(t)=[h_{ik}(t)]_{N\times K}$, where $h_{ik}(t)=\sum_{C_j=k}b_{ij}(t)$, represents the contribution of node $i$ to community $k$. Suppose node $i$ is moved from community $C_i$ to community $C_j$, then for the network $G(t)$:
\begin{equation}\label{e7}
dQ_t=Hnm(t)(i,C_j)-Hnm(t)(i,C_i)+DB(t)(i,i)
\end{equation}
Meanwhile, $Hnm(t)$ needs to be recalculated as follows:
\begin{equation}\label{e8}
\begin{split}
   Hnm(t)'(:,C_j)= & Hnm(t)(:,C_j)+DB(:,i) \\
   Hnm(t)'(:,C_i)= & Hnm(t)(:,C_i)-DB(:,i)
\end{split}
\end{equation}

In particular, given the adjacency matrix $A=(A(t),t=1,2,\cdots,T)$ and the volatility matrix $V=[V_{ij}]_{N\times N}$ of a time-varying network, the extended Louvain method for the maximization of $DQ$ as summarized as follows:

(i) First, we regard each node as one community and initialize $DQ=-\infty$.

(ii) We calculate the dynamic modularity matrix $DB(t)$, contribution matrix $Hnm(t)$ for each layer of the network, as well as the $DQ$ of the whole network.

(iii) For each node in network, we calculate $dDQ$ for all of its adjacent communities. Finding the maximum $dDQ$ that is positive, we move the node to the corresponding community and update $Hnm(t)$ for each layer of the network. We repeat this step until changing the affiliation of node to communities can not bring about the increase of $DQ$.

(iv) We reconstruct the network according to the new community structures by treating each community as one node, and repeat step (ii) and (iii) until $DQ$ does not increase.

It should be noted that the edge weight $a_{ij}(t)\in$[0,1], while the volatility $V_{ij} \in R$. In order to balance the importance of the edge weight and volatility, $V_{ij}$ is normalized as follows:
\begin{equation}\label{e9}
    v_{norm}=\frac{2}{1+exp(-\lambda v)}-1
\end{equation}
where $\lambda$ is used to regulate the variance of $v_{norm}$. Here we let the $\lambda$ and the reciprocal of the maximum of $v$ in the same order of magnitude.

\begin{figure*}[t]
	\subfloat[~]{\includegraphics[width=5.5in]{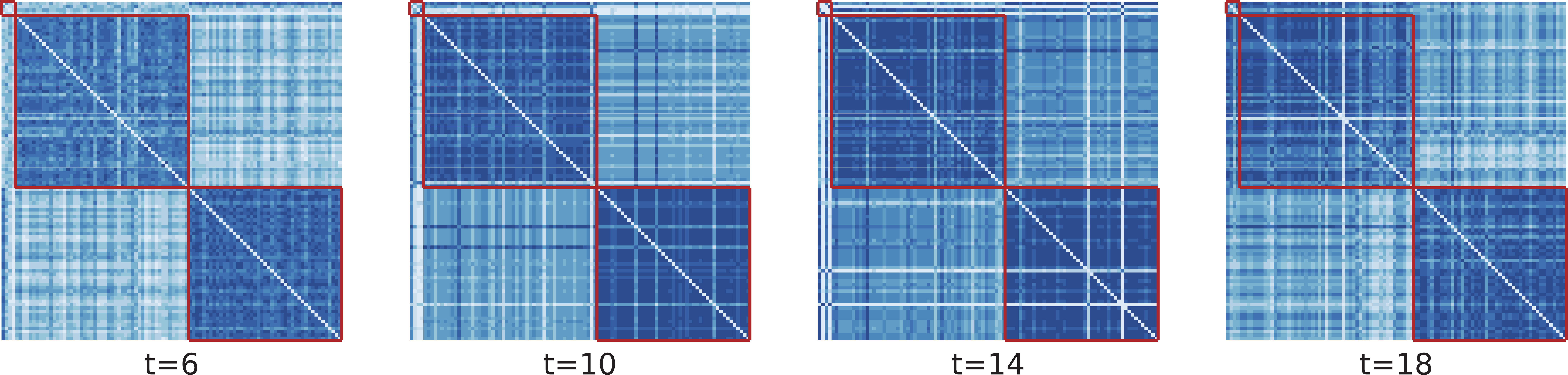}
		\label{fig_va}}
    \hfil
	\subfloat[~]{\includegraphics[width=2.0in]{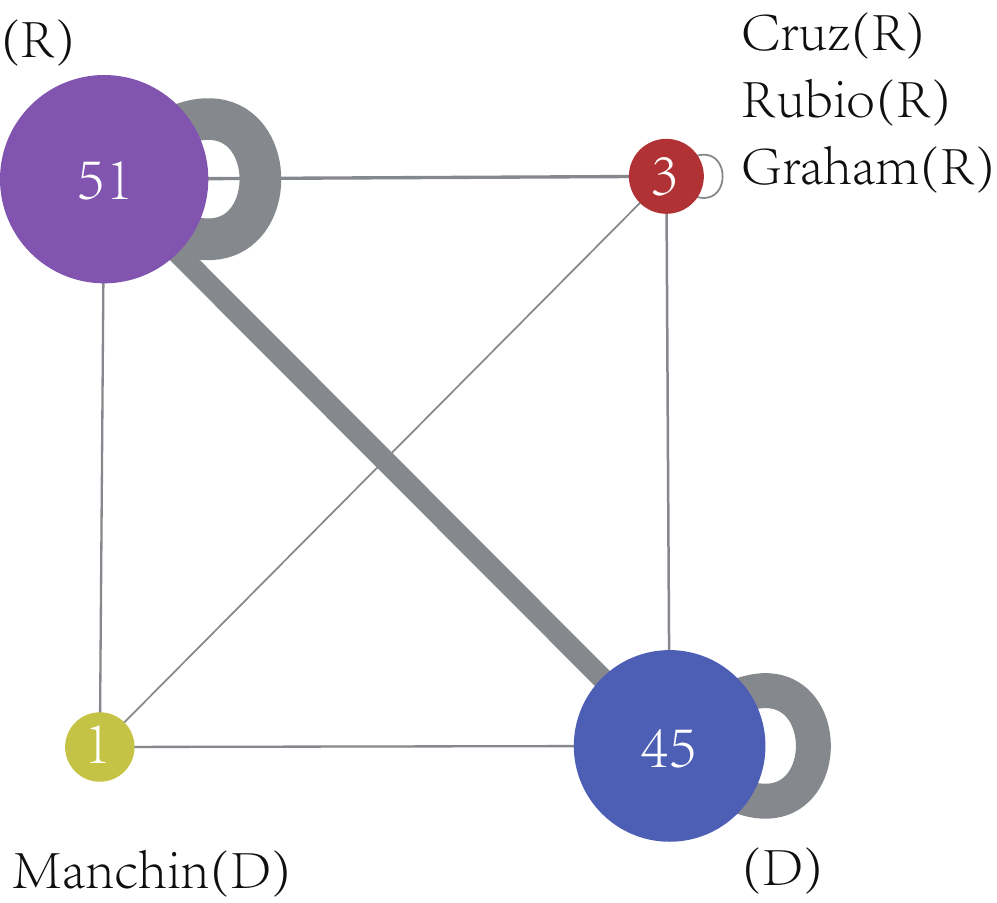}
		\label{fig_vb}}
	\hfil
    \subfloat[~]{\includegraphics[width=2.0in]{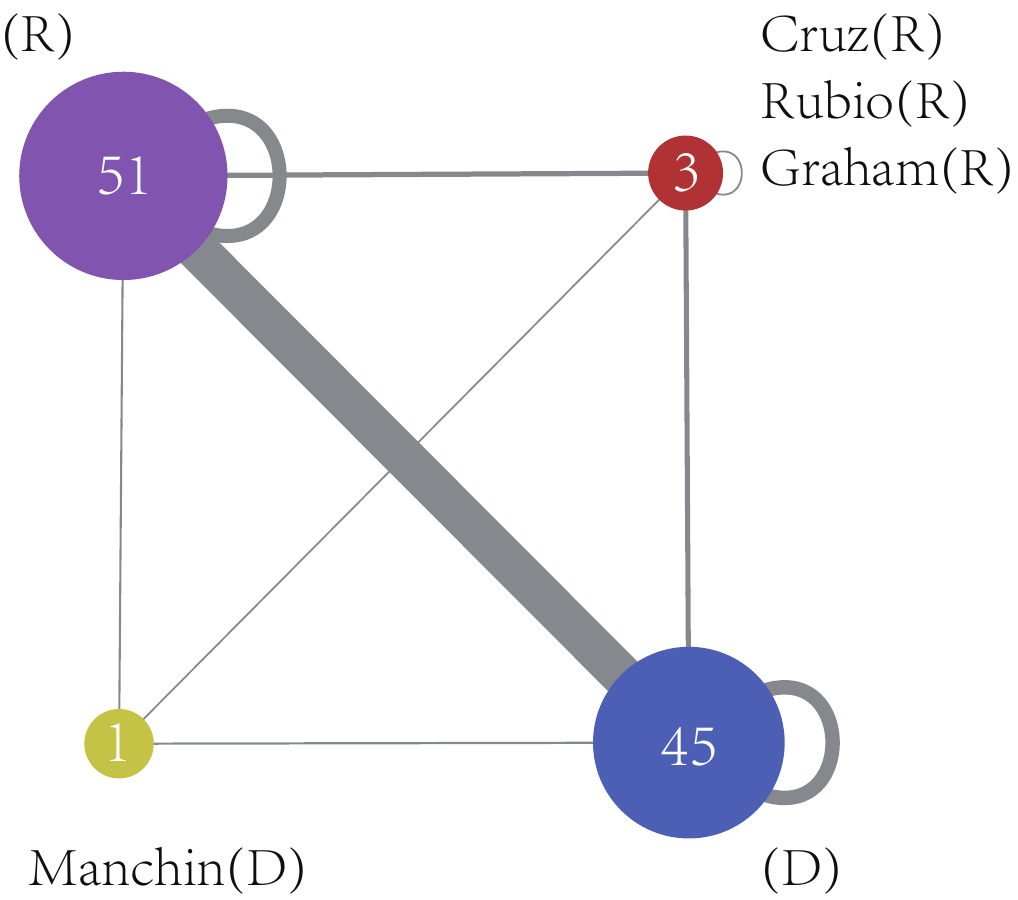}
		\label{fig_vc}}
	\caption{(Color online) Stable communities detected in the voting network by using our proposed method. (a) Adjacency matrix of the voting network at time t=6, t=10, t=14, t=18. The darkness of colors represents the value of weight. Since nodes are ordered according to the stable community structure, the diagonal blocks represent communities and are marked with the red lines. The proposed method identifies 4 communities, and the two relatively small communities are placed into the same red block. (b) The average weight of inner-community edges and inter-community edges. (c) The average volatility of inner-community edges and inter-community edges. In (b) and (c), nodes represent communities and are differentiated by node colors. The size of nodes indicates the size of communities which is equal to the number inside nodes. The thickness of edges corresponds to the value of weight in (b) and volatility in (c), respectively. Apparently, the weights of edges inside communities are greater than that of edges between communities, while the volatility of edges inside communities is smaller.}
	\label{fig_vote}
\end{figure*}

\begin{figure*}[t]
    \subfloat[~]{\includegraphics[width=5.5in]{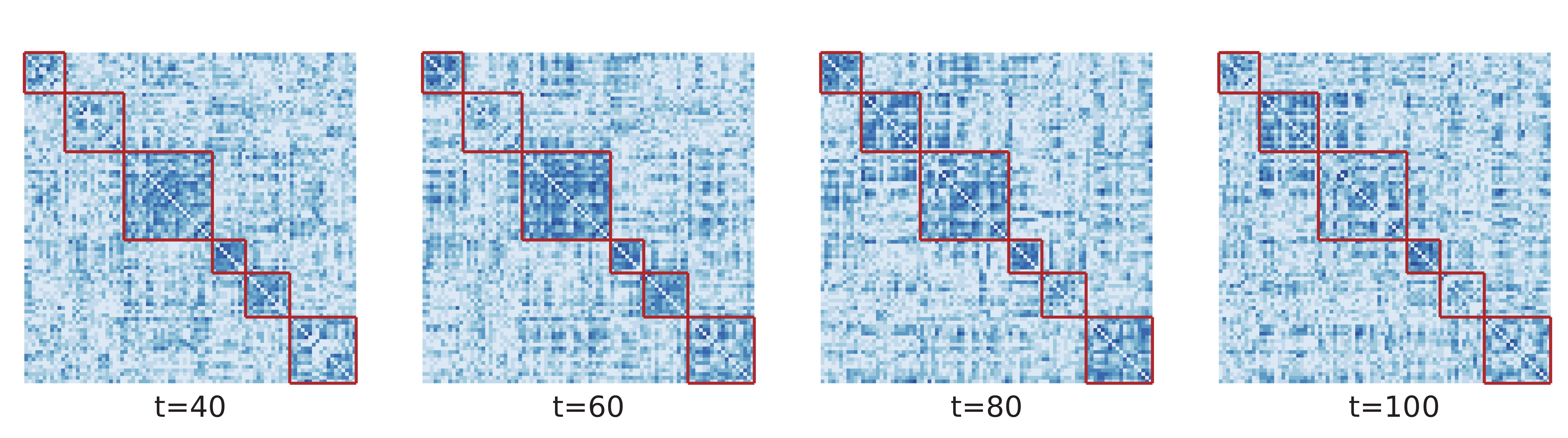}
        \label{fig_bna}}
    \hfil
	\subfloat[~]{\includegraphics[width=1.5in]{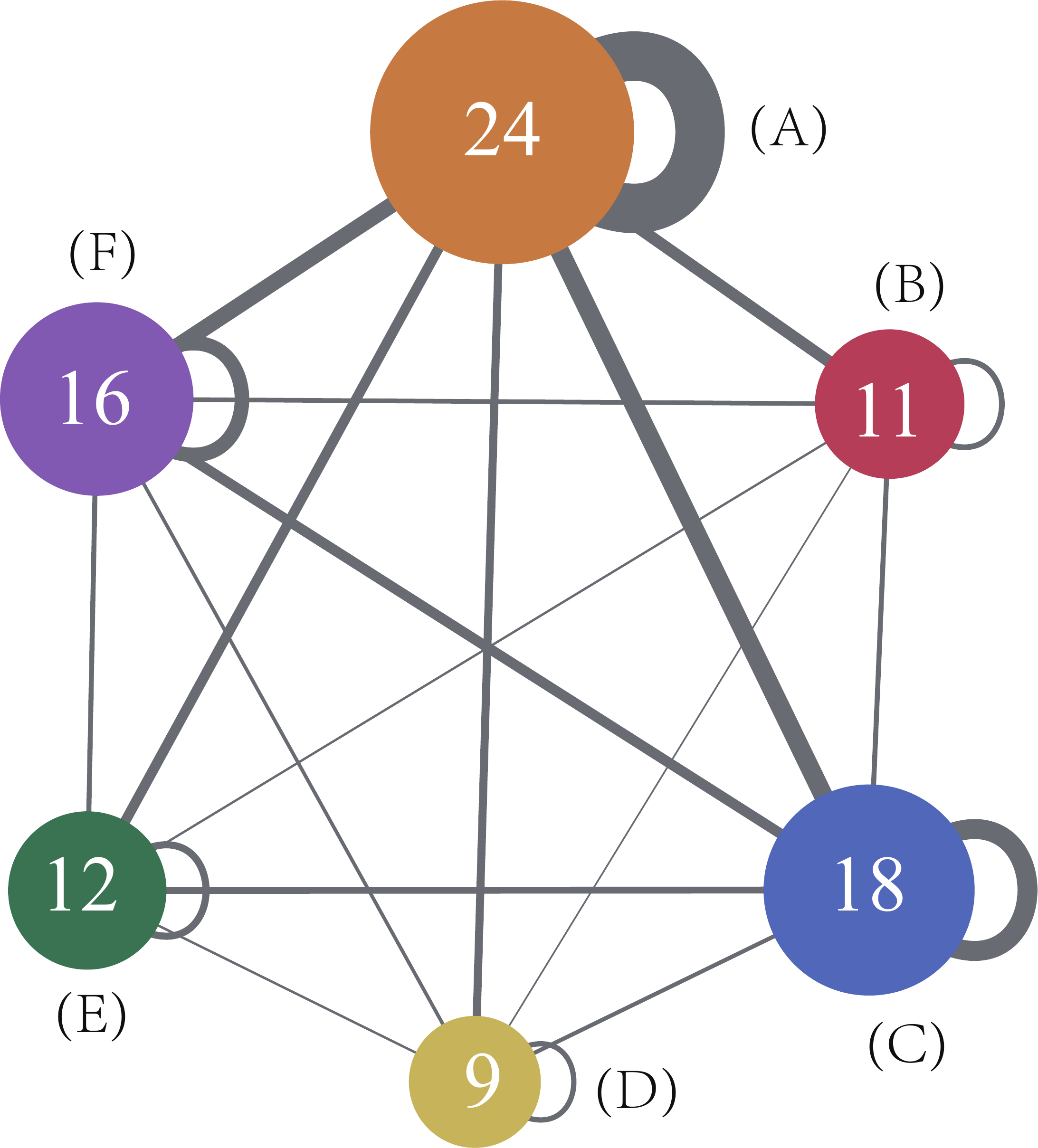}
		\label{fig_bnb}}
	\hfil
    \subfloat[~]{\includegraphics[width=1.5in]{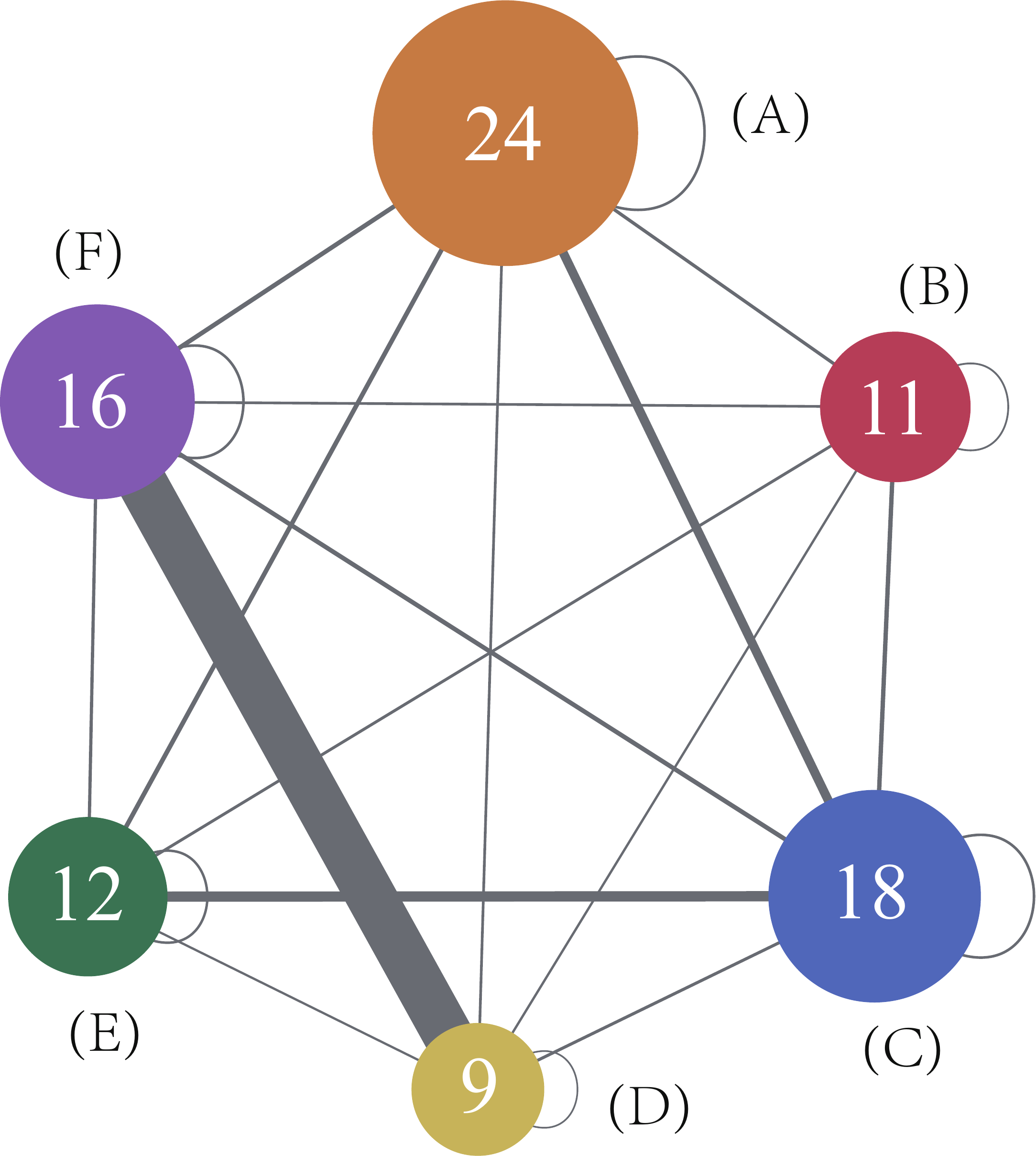}
		\label{fig_bnc}}
	\caption{(Color online) Stable communities detected by using our proposed method in the functional brain network of a randomly chosen subject. (a) Adjacency matrix of the functional brain network at time t=40, t=60, t=80, t=100. The darkness of colors represents the value of weight. There are 6 communities in total. Since nodes are ordered according to the stable community structure, the diagonal blocks represent communities and are marked with the red lines. (b) The average weight of inner-community edges and inter-community edges. (c) The average volatility of inner-community edges and inter-community edges. In (b) and (c), nodes represent communities and are differentiated by node colors. The size of nodes indicates the size of communities which is equal to the number inside nodes. The thickness of edges corresponds to the value of weight in (b) and volatility in (c), respectively. Large weight edges are inside communities, while large volatility edges are between different communities.}
	\label{fig_bc}
\end{figure*}
\subsection{\label{sec:leve25}Significance Test}
Measuring the significance of communities is an efficient way to evaluate the results of community detection\cite{Newman2004}. To verify the significance of stable community structures in real time-varying networks, we use the number of communities, $DQ$, the average weight of inner-community edges, as well as the average volatility of inter-community edges to measure the stable community structure, and adopt the hypothesis tests to compare the difference between real networks and randomized networks. Two types of randomized networks are generated by the connectional random model and temporal random model in Ref.~\cite{bassett2011dynamic}. In the connectional random model, edges inside each layer are scrambled with the distribution of degree and strength preserved,  whereas the order of layers remains unchanged. In the temporal random model, layers of the network are shuffled, while the topology of each layer remains unchanged.

For each real network, the procedure of stale community detection is repeated 100 times for the original network, and once for each of 100 connectional randomized networks and 100 temporal randomized networks, respectively. The number of communities, $DQ$, the average weight of inner-community edges and the average volatility of inter-community edges are computed based on each detection result. The Shapiro-Wilk test\cite{shapiro1965analysis} is employed to analyze the distribution type of the four variables mentioned above. If the variable obeys a normal distribution in both real networks and the randomized networks, then we utilize the independent $T$ test to perform the hypothesis test, otherwise the Mann-Whitney $U$ test \cite{mann1947test}.  The variable is considered to be significantly different between real networks and the randomized networks when the $p$ value is smaller than 0.05.

\subsection{\label{sec:leve26}Robustness Test}

A reliable community detection result should be robust against small changes of networks. In other words, the stable community structure will not become completely different when there are small perturbations in the time-varying network. To test the robustness of community structures obtained through our method, we evaluate the similarity of stable community structures in original networks and disturbed networks by the normalized mutual information (NMI)\cite{danon2005comparing}. With respects to the perturbations of networks, two kinds of perturbations are introduced including the perturbation of edge weights and the perturbation of the edge number. We could perturb edge weights as follows: first, randomly choose four edges $e_{ab}$, $e_{bc}$, $e_{cd}$ and $e_{da}$ which form a circle. Then adjust weights of these edges to a small degree, and the new weights are $\omega_{ab}-\Delta$, $\omega_{bc}+\Delta$, $\omega_{cd}-\Delta$, $\omega_{da}+\Delta$, where $\Delta$ is a random number. The strength of nodes after such processing is the same as that in the original network. As for the perturbation of edge number, we could randomly remove an edge $e_{ab}$ and connect a pair of nodes $c$ and $d$ with the same edge weight as that of edge $e_{ab}$. The total number of edges in the network does not change. The proportion of the modified edges can be considered as an indicator of the perturbation level in networks, which is termed the adjustment level $\alpha$.

The robustness test scheme is summarized as follows:

(i) For each layer of the network, we randomly choose an edge $e$, and then perturb the edge weights with a probability of 0.5 or perturb the edge number with a probability of 0.5. Repeat this step until the adjustment level $\alpha$ reaches the predefined value.

(ii) We generate $K$ perturbed networks according to step (i), and detect the stable communities for each perturbed network by using the proposed method.

(iii) We calculate the NMI of the community structures of the disturbed network and that of the real network, as well as the average of $K$ NMIs.

(iv) We change the value of the adjustment level $\alpha$ and repeat step (i)-(iii), and then we observe the change in the average of NMIs.

\section{\label{sec:leve3}Experiments on Real Networks}
In this section, we analyze two kinds of time-varying networks including the voting network and the functional brain network to demonstrate the performance of the proposed method, and compare it with the multi-layer modularity method given by Zhang $et.al.$\cite{zhang2017finding}.
\subsection{\label{sec:leve31}The Voting Network}
We build the voting network using voting records of the United States 114th Congress in 2015 and 2016\cite{TP-toolbox-web}. Each senator is denoted as a node and each layer of the network is constructed with the voting records during one month. The edge weight represents the proportion of bills where both senators share the same attitude in the month. After the records in months when the number of bills is less than five excluded, the time-varying voting network consists of 20 layers.

Our proposed method identifies $4$ stable communities in the voting network with the maximized dynamic modularity $DQ=0.1415$. In comparison, the voting network is partitioned into only 2 communities roughly by the method described in \cite{zhang2017finding} and $DQ=0.0934$, which is far less than that of our proposed method. In more detail, as shown in \reffig{fig_vote}, although edges vary with time, the weight of inner-community edges is significantly larger than that of inter-community edges, and the community structure is stable across layers. Each community either has notably strong connections between nodes inside it or is separated from other communities by significantly dynamic edges. There are two large communities which are made up of 51 republicans (denoted by R), 43 democrats and 2 independents (denoted by D), respectively. Besides, there are another one community consists of 3 republicans, $i.e.$, Cruz, Rubio and Graham, and the fourth community only contains one democrat, Manchin.  The proposed method can not only characterize the party affiliations well, but also capture the personalized behavior of subjects precisely. Although Sanders and King are independents, they are partitioned into the same community as most democrats, which is consistent with the fact that they both caucus with the Democratic Party.

To quantitatively measure the tendency of a senator to the two parties, we compare the voting records of the senator with the records of the R community and the D community. In particular, if more than two-thirds of senators in the community agree to a bill, we regard the community supports the bill. If the percentage is less than one third, we regard the community disagrees with it. Otherwise, the community takes a neural stand.  The numbers of bills where a senator stays neutral, has the same opinion as the R community, and agrees with the D community are listed in \reftab{tab_rd}. Clearly, Sanders and King lean more toward the Democratic Party, while Rubio, Graham and Cruz are almost impartial. Furthermore, Graham and Rubio often share the same opinion and maintain a strong relationship over a long period according to the reports. For example, Rubio and Graham called for stronger Russian sanctions together in July 2018 and they offered high praise of each other in 2014 and 2016. Manchin is a democrat and his votes tend to be in favour of the Democratic Party apparently. Nevertheless, Manchin is grouped into a single community, which is supported by his bipartiship and his role as a conservative Democrat.

\begin{table}[b]
\caption{\label{tab_rd}%
The distribution of voting records of some senators. Total represents the number of votes in 2015 and 2016.
}
\begin{ruledtabular}
\begin{tabular}{cccc}
\textrm{Senator} &
\textrm{Not\footnote{The number of bills where the senator stays neutral.}} &
\textrm{Rep\footnote{The number of bills where the senator has the same opinion as the R community.}} &
\textrm{Dem\footnote{The number of bills where the senator has the same opinion as the D community.}}\\
\colrule
\textrm{Sanders} & 448 & 9 & 45  \\
\textrm{King} & 428 & 15 &  59  \\
\textrm{Graham} & 452 & 34 & 34       \\
\textrm{Rubio} & 457 & 25 & 20   \\
\textrm{Cruz} & 449 & 25 & 28 \\
\textrm{Manchin} & 419 & 18 & 65  \\
\end{tabular}
\end{ruledtabular}
\end{table}

\begin{figure*}[t]
	\centering
	\subfloat[~]{\includegraphics[width=1.6in]{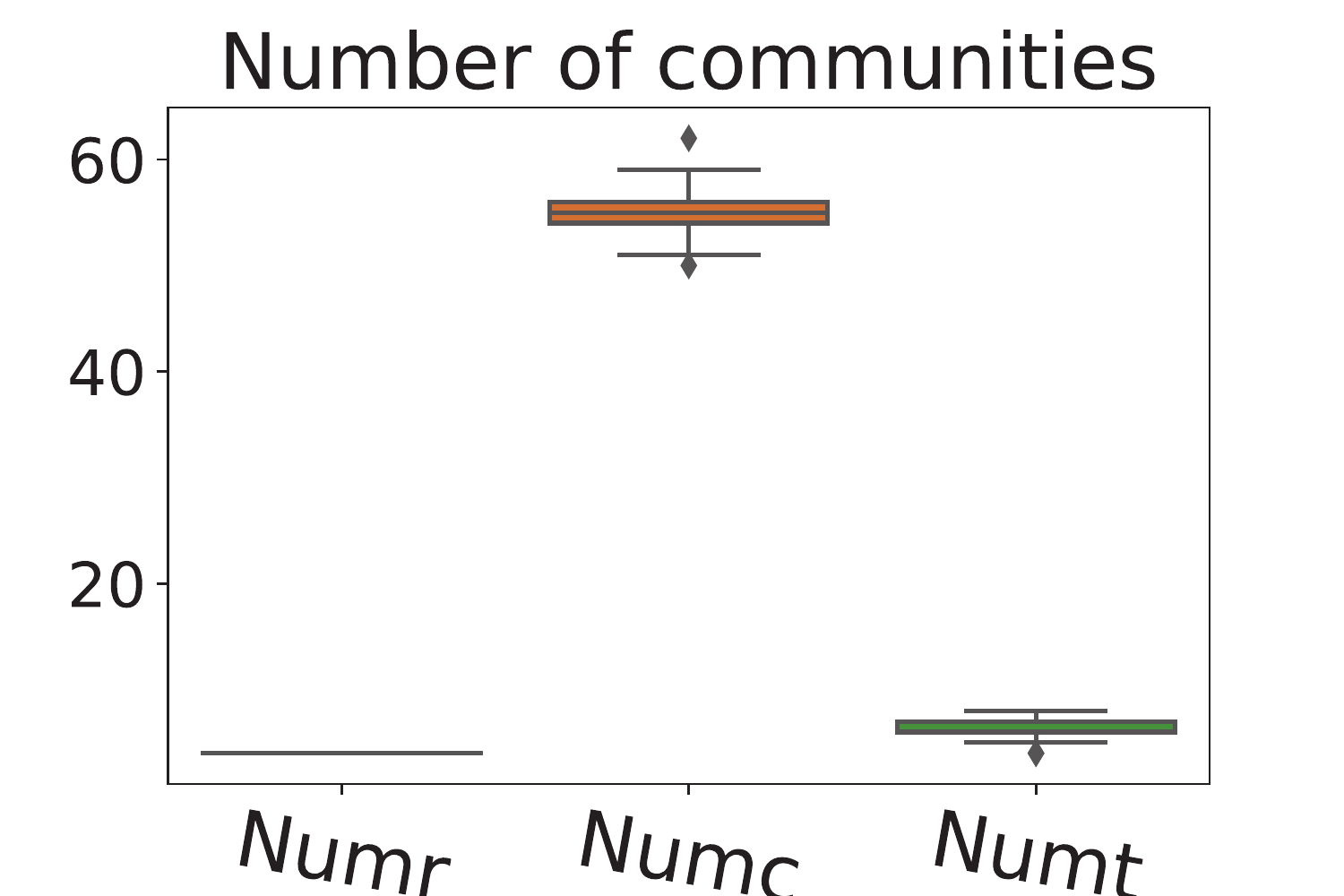}
		\label{fig_rnQ_vote}}
	\hfil
	\subfloat[~]{\includegraphics[width=1.6in]{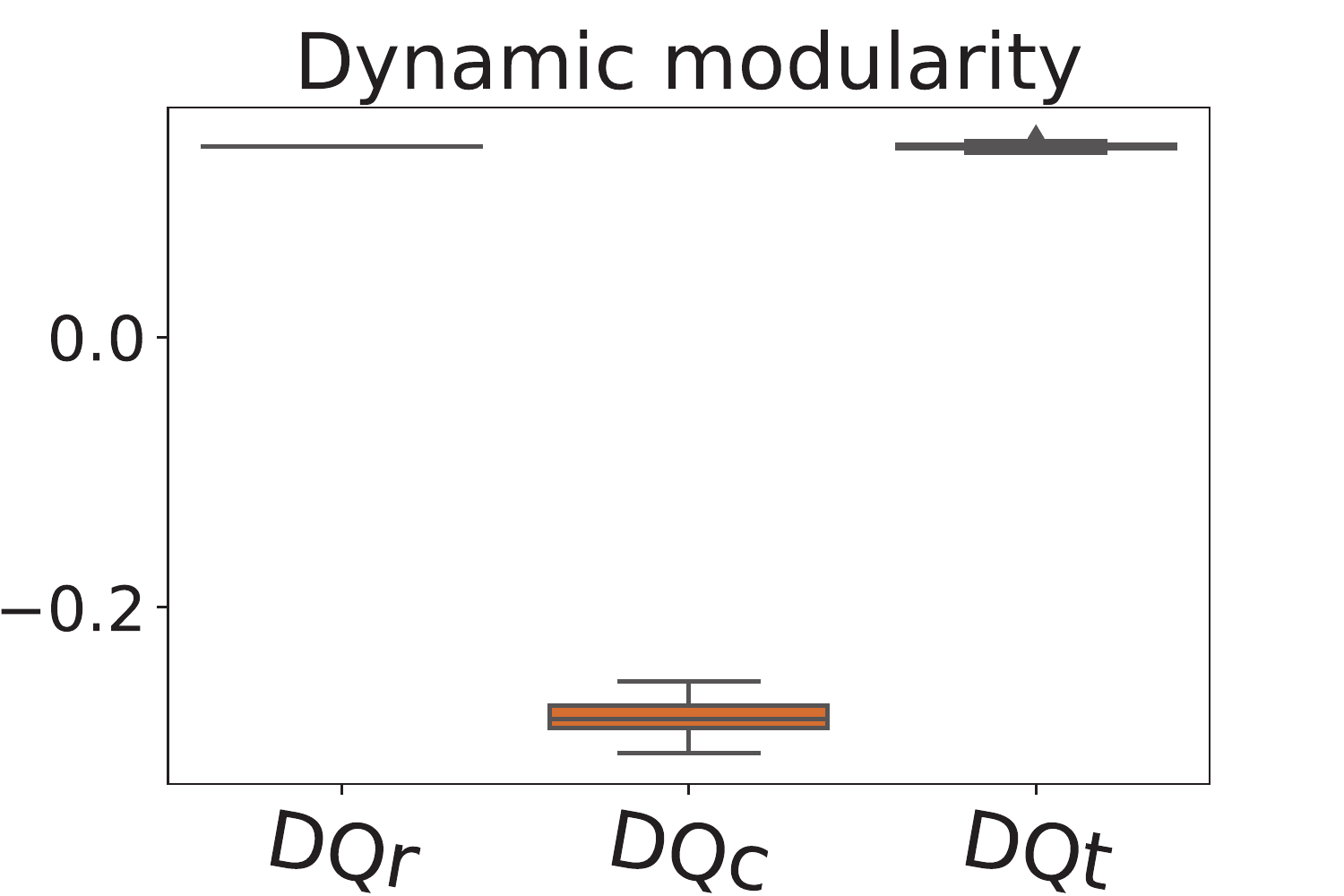}
		\label{fig_rnW_vote}}
	\hfil
    \subfloat[~]{\includegraphics[width=1.6in]{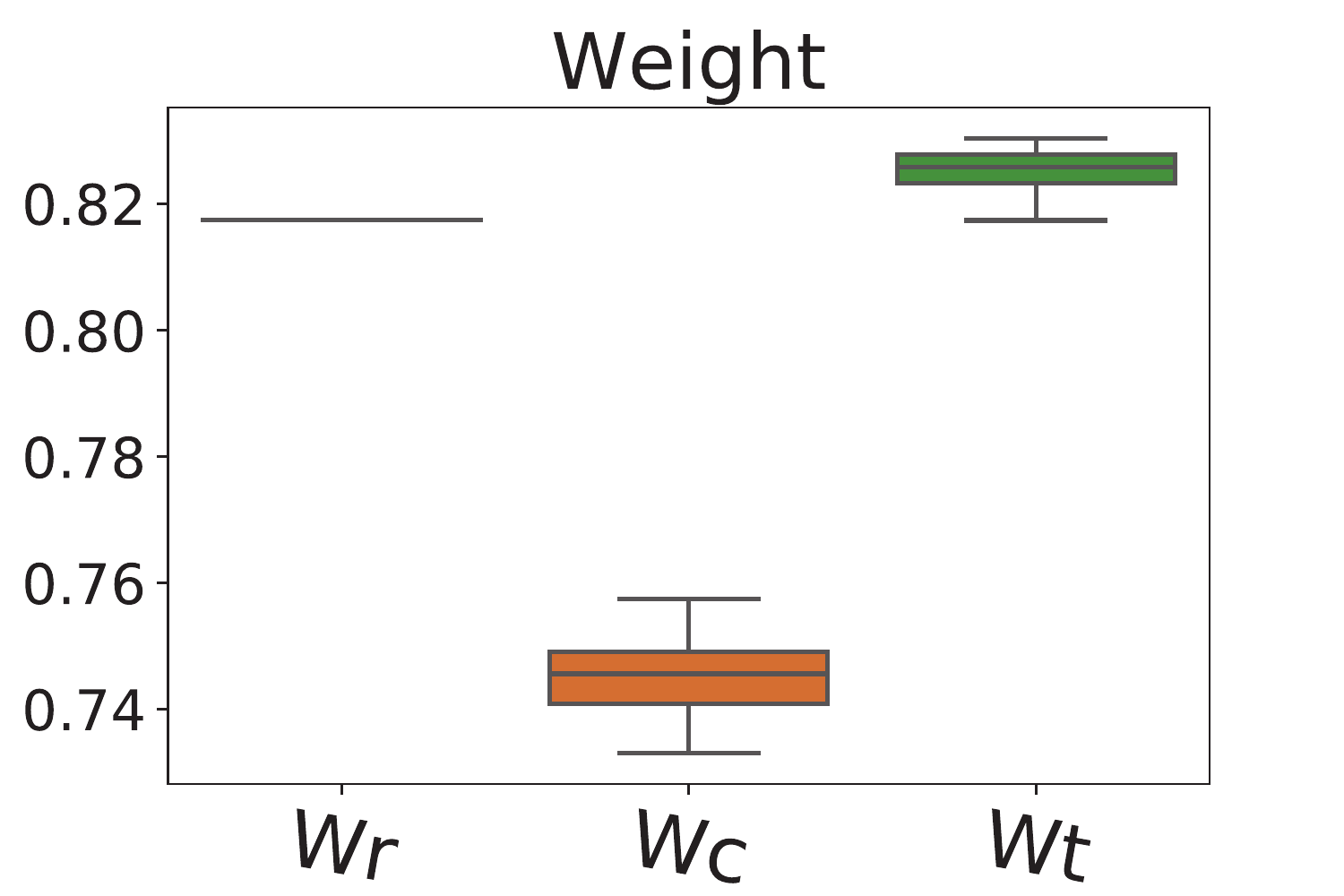}
		\label{fig_rnV_vote}}
    \hfil
    \subfloat[~]{\includegraphics[width=1.6in]{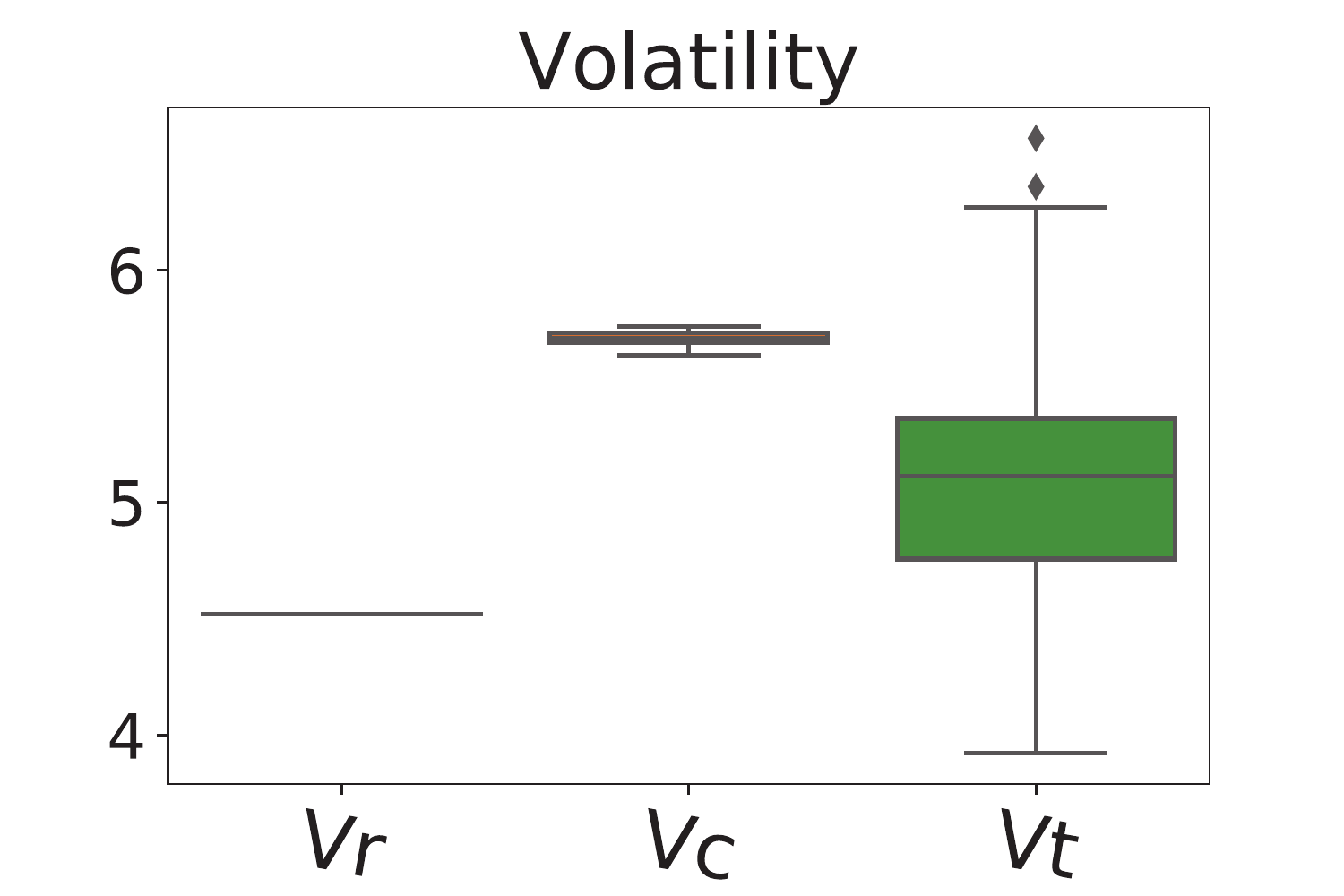}
		\label{fig_rnN_vote}}
    \hfil
	\subfloat[~]{\includegraphics[width=1.6in]{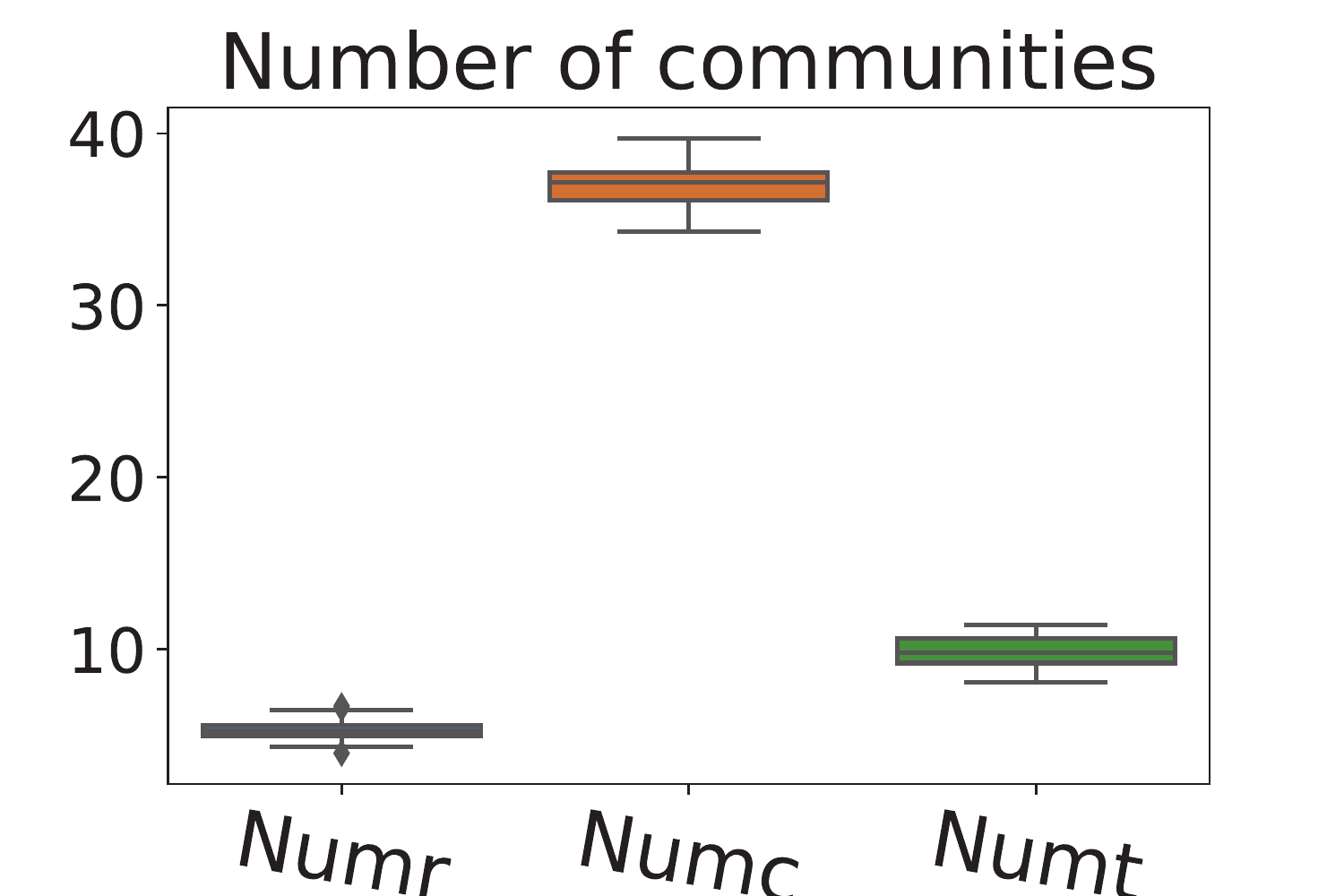}
		\label{fig_rnQ}}
	\hfil
	\subfloat[~]{\includegraphics[width=1.6in]{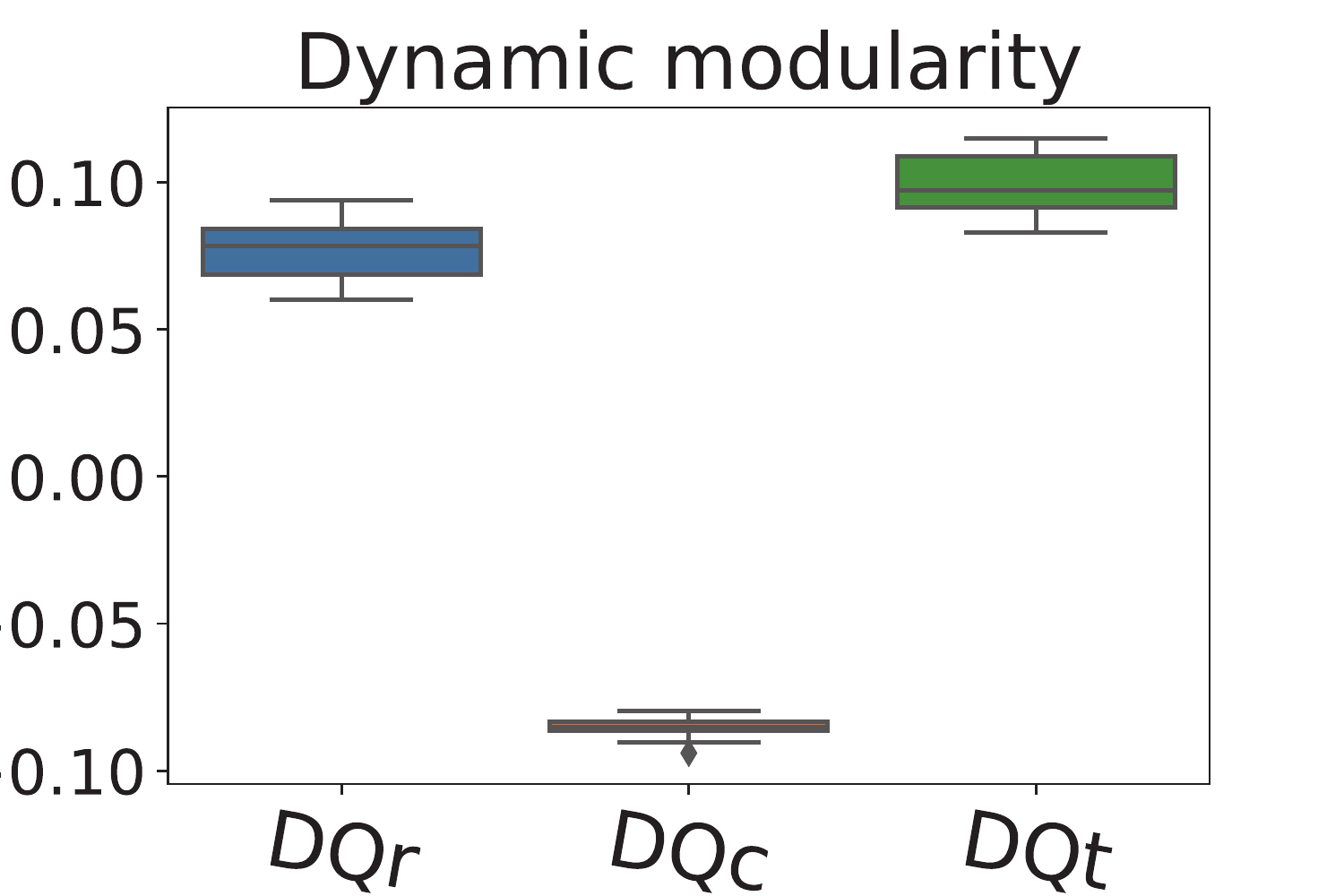}
		\label{fig_rnW}}
	\hfil
    \subfloat[~]{\includegraphics[width=1.6in]{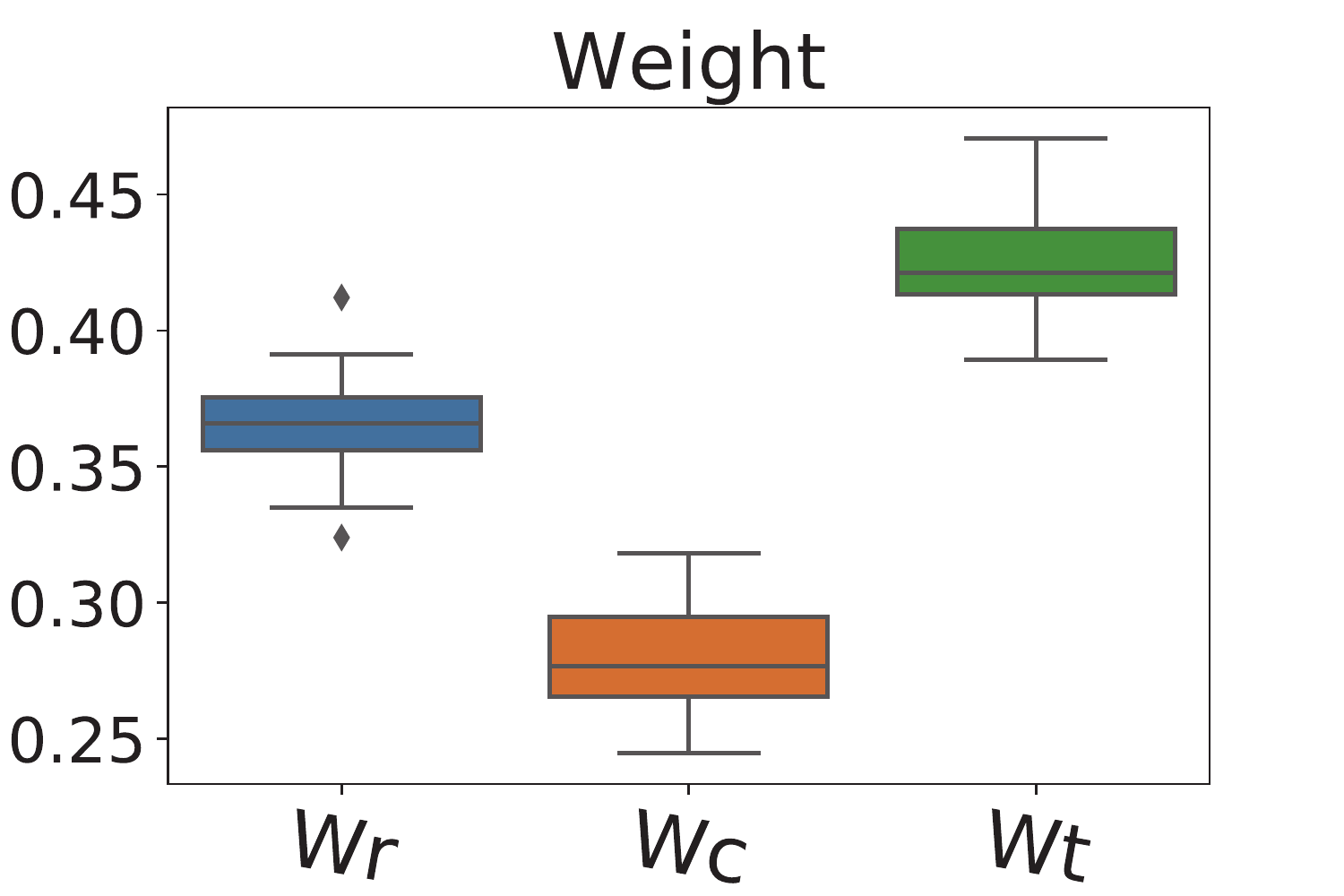}
		\label{fig_rnV}}
    \hfil
    \subfloat[~]{\includegraphics[width=1.6in]{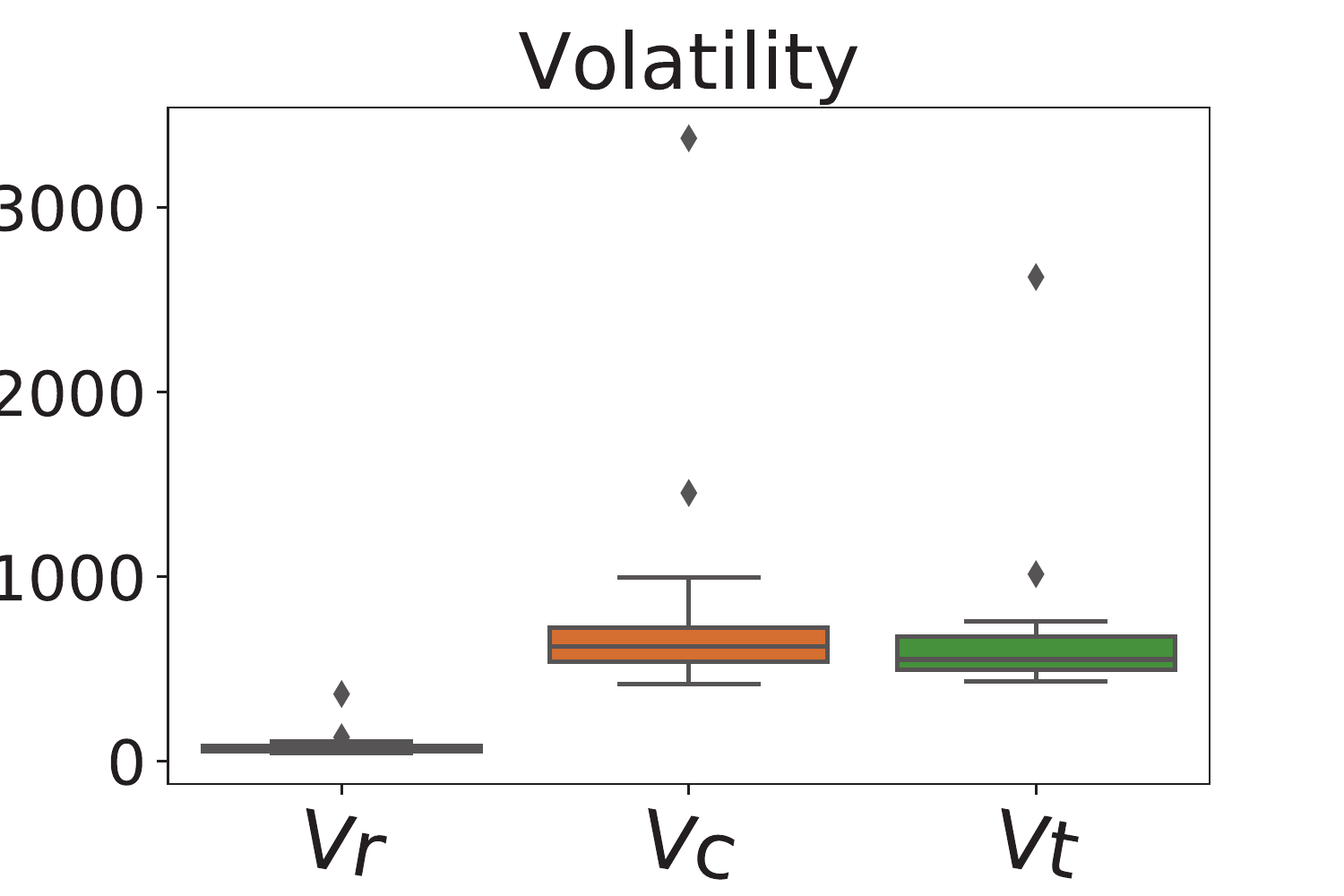}
		\label{fig_rnN}}
	\caption{(Color online) Box plots of number of communities, dynamic modularity, the average weight of inner-community edges and the average volatility of inter-community edges in real networks, connectional randomized networks and temporal randomized networks, which are denoted by the subscripts r, c, t, respectively. The three types of networks are discriminated by bar colors. The first row (a)-(d) shows the difference of the four measures between the voting network and 100 randomized networks, and the second row (e)-(h) shows the difference of the average of the four measures among 100 detection results between functional brain networks and randomized networks of 26 subjects. All of the four measures in real networks are significantly different from that in both types of randomized networks.  }
	\label{fig_cop}
\end{figure*}

\subsection{\label{sec:leve32}The Functional Brain Network}
Now we utilize the resting state functional MRI (fMRI) datasets of Attention Deficit Hyperactivity Disorder 200 Sample (ADHD-200) data, acquired from the R-fMRI Maps Project Data Release 170717\cite{Rfmriweb}. Here we use the dataset obtained from New York University (referred as the NYU dataset) to construct functional brain networks. All data is acquired at 3T Siemens Scanner. Detailed scan parameters and preprocessing steps refer to Ref.~\cite{adhdweb}. It is worthy noting that the fMRI data from the subjects with large head motions, $i.e.$, maximal motion between volumes in each direction $>3mm$ and rotation about each axis $>3^\circ$ \cite{cheng2015voxel} or the percent of framewise displacement $>0.5$ more than 10\%, are excluded to ensure the data quality. Finally we analyze the resting-state fMRI time series from 26 healthy controls in the NYU dataset.

According to the Automated Anatomical Labeling (AAL) template\cite{tzourio2002automated}, the brain is divided into 90 regions, which are defined as nodes. The time-varying functional brain networks are built with the time window shifting method. When the length of time window ($L$) is equal to the reciprocal of the minimum frequency in the signal, $i.e.$, $L=\frac{1}{f_{min}}$, the impact resulting from the window shifting is the least\cite{leonardi2015spurious}. Based on the rule of thumb, since $f_{min}$ is equal to $0.01Hz$ in our data, we set $L=50$TR (repetition time, TR=2000ms) and the step is 1TR. Then all of the fMRI time series of one subject are partitioned into 126 segments. In each time window, the Pearson correlation of the fMRI time series of two brain regions are the weight of the edge linking the corresponding nodes. For one subject, the time-varying functional brain network can be represented as 126 adjacency matrices $A_{90\times90}$.

Our proposed method is applied to each time-varying functional network of 26 subjects. Among all subjects, the average number of stable communities is $5.30$ and the average of maximum $DQ$ is $0.078$. Meanwhile, we use the method in \cite{zhang2017finding} to recognize communities in time-varying functional brain networks of the 26 subjects. The results suggest that the average number of communities is $4.2$, and the average of maximum $DQ$ is $0.061$, which is less than that of community detection results given by our proposed method. As an example, \reffig{fig_bc} shows the stable communities obtained with our method of one subject who is randomly chosen. The 90 nodes of the subject are divided into 6 communities. Although the community may be not significant enough at certain time (e.g. the top-left community at t=100 as represented in \subreffig{fig_bc}{fig_bna}), it can be identified successfully by our method which incorporates all the information across the whole period. Meanwhile, large weight edges are inside communities and large volatility edges are between different communities.

\begin{figure}[t]
	\centering
    \subfloat[~]{\includegraphics[width=2.1in]{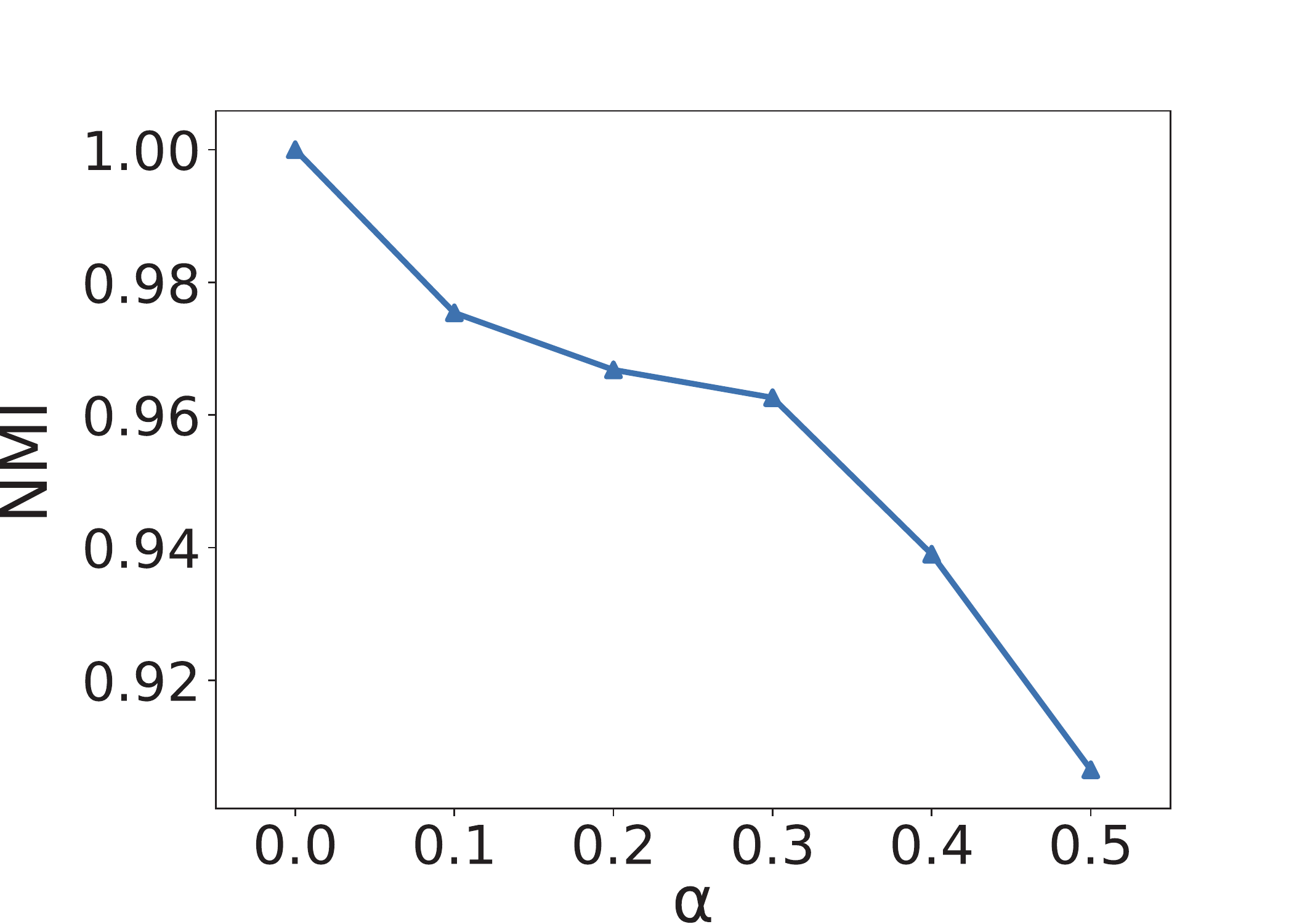}
		\label{fig_rb_vote}}
    \hfil
    \subfloat[~]{\includegraphics[width=2.1in]{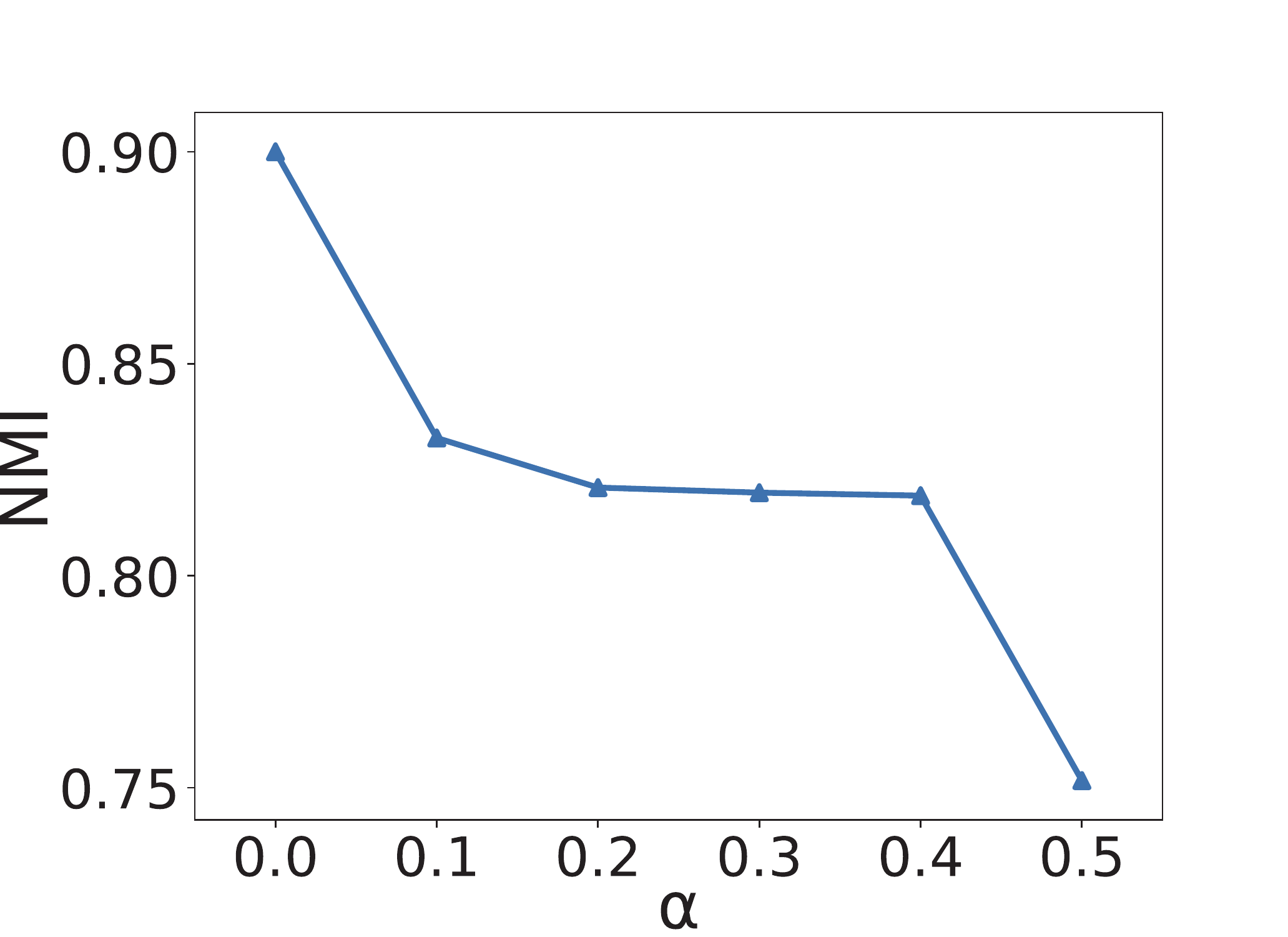}
		\label{fig_rb_brain}}
	\caption{(Color online) The average of NMI between community structures in the real network and 10 perturbed network at different adjustment level $\alpha$. (a) The result in the voting network. (b) The result in the functional brain network of a subject who is randomly chosen. The value of NMI decreases with $\alpha$ increasing in both cases.}
	\label{fig_fn}
\end{figure}

\subsection{\label{sec:leve33}Significance Test Results}
We perform significance tests for stable communities of the voting network and the time-varying functional brain networks. Regarding the voting network, we compare the difference of stable communities between the real network and 100 randomized networks (see \reffig{fig_cop}\subref{fig_rnQ_vote}-\subref{fig_rnN_vote}) . Regarding the functional brain network, for each subject, we calculate the average of a measure among 100 detection results of real networks, connectional randomized networks and temporal randomized networks, respectively. We compare the difference of the average of four measures between the real networks and randomized networks of 26 subjects.(see \reffig{fig_cop}\subref{fig_rnQ}-\subref{fig_rnN}). The results of hypothesis tests show that the p-value of all tests are far less than $0.0001$ with only the test of $DQ$ in the voting network and temporal randomized networks at $p=0.25$, $i.e.$, the number of communities, $DQ$, the average weights of inner-community edges and the average volatility of inter-community edges between real networks and randomized networks are significantly different, except $DQ$ in the real network and temporal randomized networks which may be related with the relatively minor difference, and contingency caused by the small number of real networks.

As shown in \reffig{fig_cop}, the differences of stable communities between real networks and randomized networks in the case of the voting network are similar to that in the case of functional brain networks. Compared to the original network, the nodes in both types of randomized networks are separated into more communities by the edges with much higher volatility. The $DQ$ and average weight of inner-community edges in real networks are obviously larger than those in connectional randomized networks, however, smaller than that in temporal randomized networks.

\subsection{\label{sec:leve34}Robustness Test Results}
We test the robustness of stable communities in the voting network and the time-varying functional brain networks as well. \reffig{fig_fn} shows the average NMI between the stable community structures in real networks and $10$ perturbed networks at different adjustment level $\alpha$. No matter in the voting network or the functional brain network, the value of NMI decreases with $\alpha$ increasing, and the larger the adjustment level, the greater the degree of changes in the network structure. The NMI remains higher than 0.9 in the voting network, and higher than 0.75 in the functional brain network, when $\alpha$ is less than 0.5, which indicate that the community structure given by our method is robust.
\begin{table}[tb]
\caption{\label{tab_d}Demographic information for the NYU dataset and the PKU dataset}
\begin{ruledtabular}
\begin{tabular}{cccccc}
\textrm{Site} & \textrm{Group} & \textrm{Num} & \textrm{Age} & \textrm{Sex(M/F)} & \textrm{ADHD index} \\
\colrule
\multirow{2}*{\textrm{NYU}} & \textrm{ADHD} & 30 & $10.8\pm2.6$ & 20/10 & $72.5\pm 10.7$\\
~ & \textrm{Control} & 26 & $12.1\pm 3.1$ & 8/18 & $45.1\pm5.3$ \\
\cline{1-6}
\multirow{2}*{\textrm{PKU}} & \textrm{ADHD} & 88 & $12.1\pm2.0$ & 77/11 & $49.4\pm 7.9$\\
~ & \textrm{Control} & 133 & $11.5\pm 1.9$ & 76/57 & $29.4\pm6.5$ \\
\end{tabular}
\end{ruledtabular}	
\end{table}

\begin{table}[tb]
\caption{\label{tab_f}Features of time-varying functional brain networks for training the classification model}
\begin{ruledtabular}
\begin{tabular}{lllr}
\textrm{Category of Features} & \multicolumn{2}{c}{\textrm{Features}}  \\
\colrule
\textrm{Global property} & \textrm{DQ} & \multicolumn{2}{l}{\textrm{Qnum}} \\
\colrule
\textrm{Edge weight} & \textrm{W\_inner}\footnote{The suffix \_inner denotes the average of the index among inner-community edges.} & \multicolumn{2}{l}{ \textrm{W\_inter}\footnote{The suffix \_inter denotes the average of the index among inter-community edges.}} \\
\colrule
\multirow{3}*{\textrm{Volatility}}  & \textrm{V\_inner} & \multicolumn{2}{l}{ V\_inter} \\
~ &  \textrm{Vratio\_inner} & \multicolumn{2}{l}{\textrm{Vratio\_inter}} \\
~ &  \textrm{Vpratio\_inner} & \multicolumn{2}{l}{\textrm{Vpratio\_inter}} \\
\colrule
\textrm{Entropy of volatility} & HV\_inner & HV\_inter & HV \\
\end{tabular}
\end{ruledtabular}	
\end{table}

\begin{table*}[htb]
\caption{\label{tab_c}The top four classifiers for the NYU dataset and the PKU dataset.}
\begin{ruledtabular}
\begin{tabular}{ccccccccc}
Datasets & \multicolumn{3}{c}{\textrm{Features used for training the classifiers}} & \textrm{Accuracy} & \textrm{Specificity} & \textrm{Sensitivity} & \textrm{F1} & \textrm{AUC}\\
\colrule
\multirow{4}*{\textrm{NYU}} & \textrm{Vpratio\_inner}& & \textrm{V\_inter} & 0.7857 & 0.7768 & 0.7976 & 0.7626 & 0.8601\\
~ & \textrm{Vpratio\_inner} & \textrm{Hv\_inter} & \textrm{Hv\_inner} & 0.6786 & 0.625 & 0.7381 & 0.6896 & 0.8839\\
~ & \textbf{Vpratio\_inner} & \textbf{HV} & \textbf{Hv\_inner}\footnote{The best classifier for the NYU dataset} & 0.8036 & 0.8214 & 0.8036 & 0.7954 & 0.8810  \\
~ & \textrm{W\_inter} & \textrm{HV} & \textrm{Vraio\_inner} & 0.6786 & 0.7946 & 0.5595 & 0.6276 & 0.8780  \\
\cline{1-9}
\multirow{4}*{\textrm{PKU}} & \textbf{HV} & \textbf{Hv\_inter} & \textbf{Hv\_inner}\footnote{The best classifier for the PKU dataset} & 0.6244 & 0.6760 & 0.5466 & 0.5234 & 0.6539\\
~ & \textrm{HV} & \textrm{Hv\_inter} & \textrm{Vratio\_inner} & 0.5973 & 0.6754 & 0.4791 & 0.4751 & 0.6461\\
~ & \textrm{V\_inter} & \textrm{Hv\_inter} & \textrm{Vratio\_inter} & 0.6154 & 0.6846 & 0.5108 & 0.5050 & 0.6380 \\
~ & \textrm{HV} & & \textrm{W\_inter} & 0.6063 & 0.6378 & 0.5554 & 0.5250 & 0.6507  \\
\end{tabular}
\end{ruledtabular}	
\end{table*}

\section{\label{sec:leve4}Classification of ADHD patients and Healthy controls}
Attention Deficit Hyperactivity Disorder (ADHD) is one of the most commonly diagnosed mental disorders of children. Recent brain imaging studies have demonstrated that the underlying neural mechanisms of the ADHD are involved in multiple functional deficits in the brain. Since communities of the functional brain networks have a close relationship with the function of brain, here we use our proposed method to identify and characterize the stable community structures of the ADHD patients. Furthermore, we utilize the difference between patients and healthy controls to classify the two groups.

Besides the NYU dataset which has been used in section \ref{sec:leve3}, we also employ the dataset obtained from Peking University (referred as the PKU dataset). With the same preprocessing steps and excluding criteria, we eventually analyze the resting-state fMRI time series from 56 subjects (healthy controls(HC)=26, ADHD=30) in the NYU dataset and 221 subjects (HC=133,ADHD=88) in the PKU dataset. The demographic characteristics for each dataset are given in \reftab{tab_d}.

We employ three types of indices to measure the characteristics of stable communities for each subject: the global property including $DQ$ and the number of communities, the volatility related indices, as well as the edge weights related indices. \reftab{tab_f} lists all of the indices that we use. With these indices as features, we employ the supporting vector machine(SVM) to classify subjects. In order to evaluate the effect of different indices, all of the combinations of two indices and, combinations of three indices are used to train classifiers, respectively. We use 10-fold cross-validation method to evaluate the quality of classifiers through indices such as accuracy, sensitivity, specificity, F1 and AUC.

\begin{figure}[t]
	\centering
	\subfloat[~]{\includegraphics[width=2.4in]{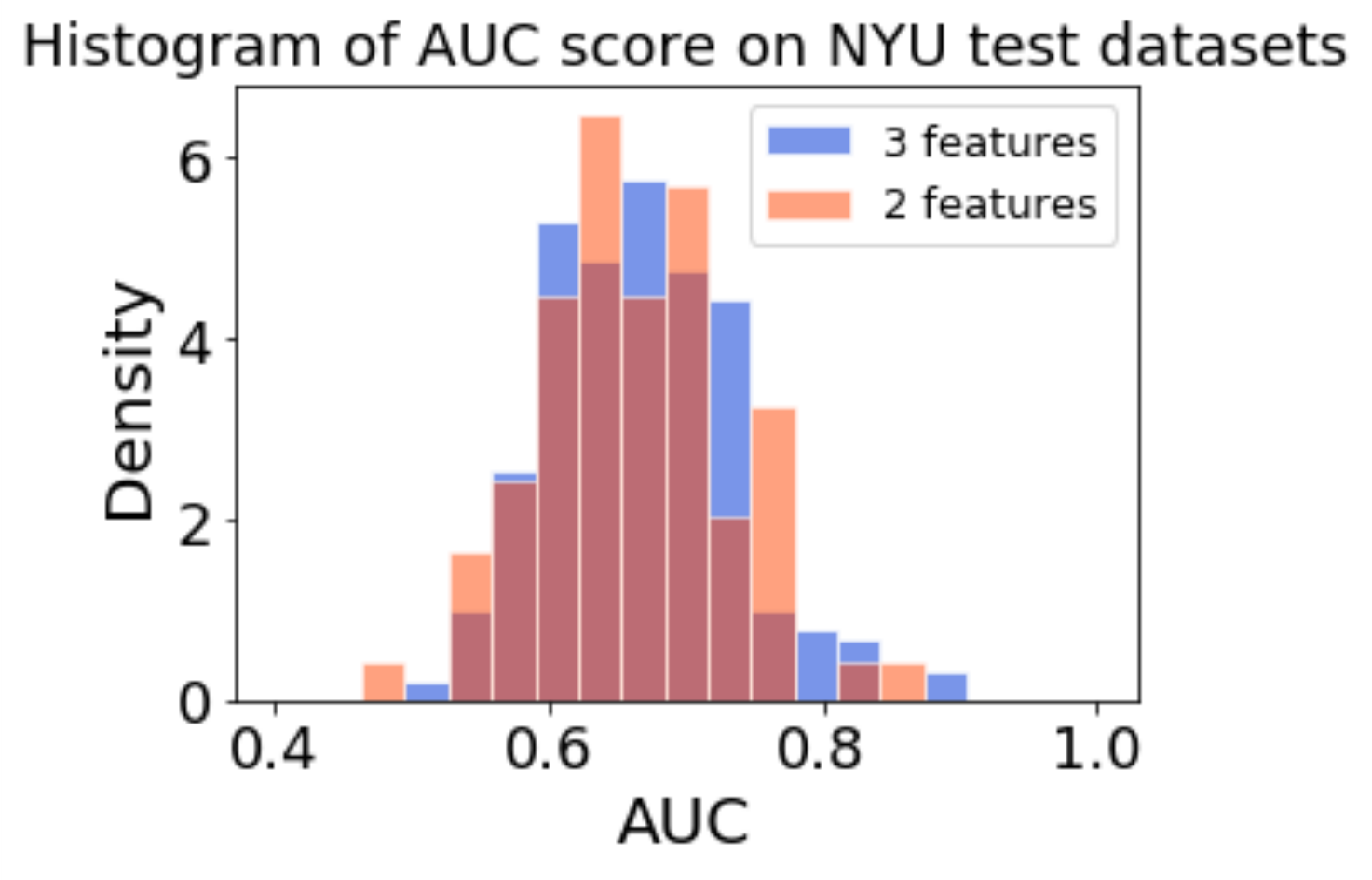}
		\label{fig_a}}
	\hfil
    \subfloat[~]{\includegraphics[width=2.3in]{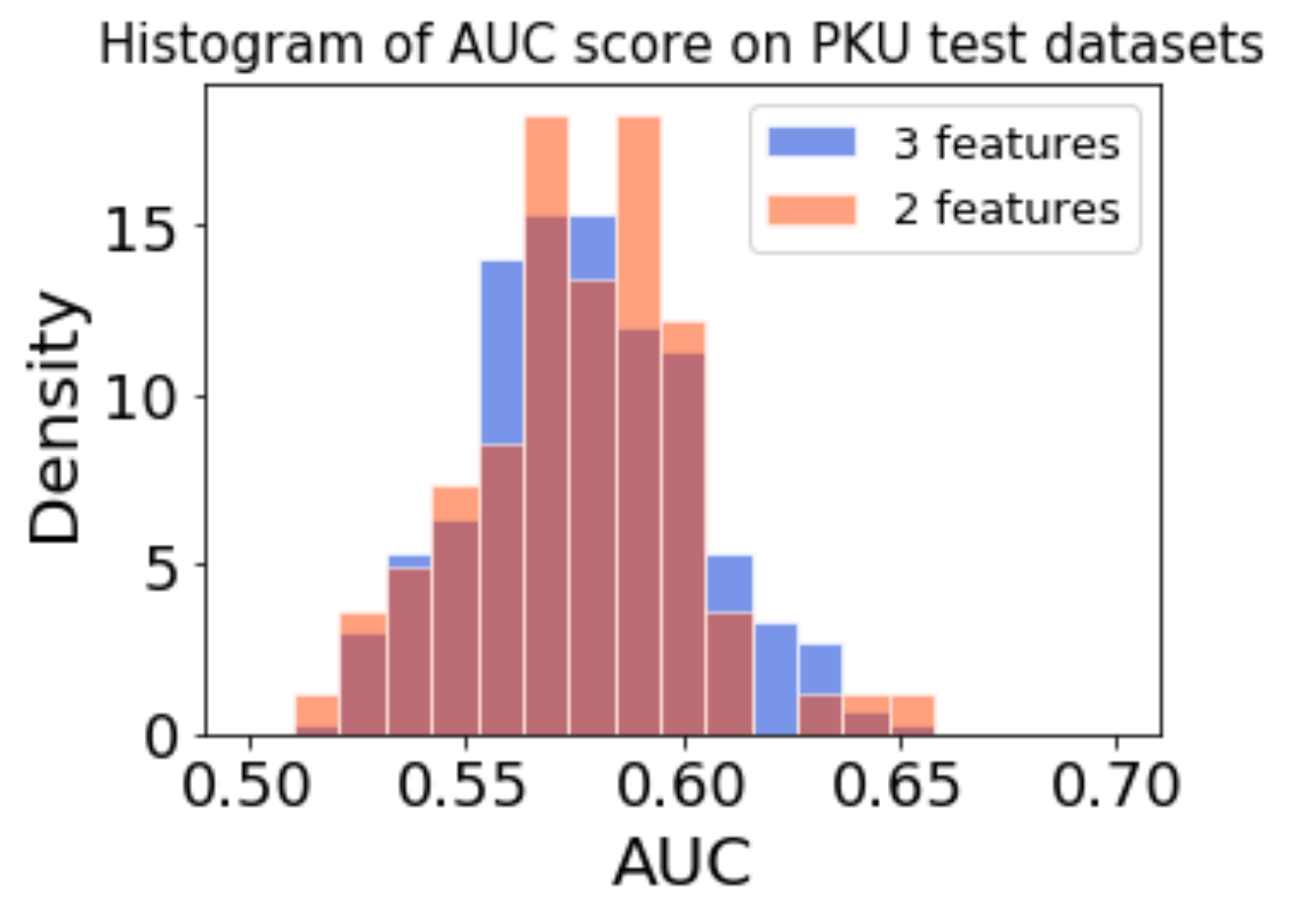}
		\label{fig_b}}
	\caption{(Color online) Histograms of the AUC score when the classifiers are trained with the combinations of two features and three features. (a) The histogram of the AUC on the NYU dataset. (b) The histogram of the AUC on the PKU dataset. In both datasets, only a few combinations of features can achieve high accuracy of classification.}
	\label{fig_c}
\end{figure}

Different combinations of indices bring different classification effect (see \reffig{fig_c}). We can see that the distribution plot of the AUC are almost the same whatever the features are the combination of two or three indices, while only a few classifiers can reach the high AUC. In the NYU dataset, four classifiers have a AUC higher than 0.85 (see the upper part of \reftab{tab_c}), and the features used for training these classifiers involve the relative volatility and the entropy of volatility. Similarly, the four classifiers with the highest AUC in the PKU dataset are trained by the relative volatility and the entropy of volatility (see the bottom part of \reftab{tab_c}). Compared to the healthy controls, such results suggest that the ADHD patients have the abnormalities on the relative volatility and the heterogeneity of volatility. Meanwhile, the effective characterization of the relative volatility and the entropy of volatility for the dynamic properties of functional brain networks demonstrates that it is highly reasonable to introduce the volatility concept to the definition of $DQ$.

In the NYU dataset, the best classifier is trained by the features including Vpratio\_inner, HV and HV\_inner, whose accuracy, sensitivity, specificity, and F1 are all ranked the first among the four classifiers given in \reftab{tab_c}, whose accuracy can reach $80.36\%$. In the PKU dataset, the best classifier is trained by Hv\_inter, HV and Hv\_inner, whose accuracy is $62.44\%$. In contrast, as the reported results of the ADHD-200 competition, the highest accuracy are only $56\%$ on the NYU dataset and $58\%$ on the PKU dataset\cite{nunez2015fadr}. Note that our method deploys a much smaller number of features and achieves the better classification performance.

\section{CONCLUSION}
In this paper, we have developed a framework to detect temporal stable communities in time-varying networks by maximizing the newly defined dynamic modularity. We conclude that dynamic modularity explicitly characterizes the temporal changes of communities through volatility, which is critical to capture the time-varying significance of real-world networking systems such as the voting networks and functional brain networks. This method can be directly used for various other tasks. The consideration of the correlation of temporal changes between different edges, and the generalization for tracking the changes of communities over time deserve more efforts to explore in future.

\begin{acknowledgments}
This work was partly supported by the National Natural Science Foundation of China (No. 71731004, No. 61603097), the National Natural Science Fund for Distinguished Young Scholar of China (No. 61425019), and Natural Science Foundation of Shanghai (No. 16ZR1446400).
\end{acknowledgments}
\bibliography{braincommu}

\begin{thebibliography}{43}%
\makeatletter
\providecommand \@ifxundefined [1]{%
 \@ifx{#1\undefined}
}%
\providecommand \@ifnum [1]{%
 \ifnum #1\expandafter \@firstoftwo
 \else \expandafter \@secondoftwo
 \fi
}%
\providecommand \@ifx [1]{%
 \ifx #1\expandafter \@firstoftwo
 \else \expandafter \@secondoftwo
 \fi
}%
\providecommand \natexlab [1]{#1}%
\providecommand \enquote  [1]{``#1''}%
\providecommand \bibnamefont  [1]{#1}%
\providecommand \bibfnamefont [1]{#1}%
\providecommand \citenamefont [1]{#1}%
\providecommand \href@noop [0]{\@secondoftwo}%
\providecommand \href [0]{\begingroup \@sanitize@url \@href}%
\providecommand \@href[1]{\@@startlink{#1}\@@href}%
\providecommand \@@href[1]{\endgroup#1\@@endlink}%
\providecommand \@sanitize@url [0]{\catcode `\\12\catcode `\$12\catcode
  `\&12\catcode `\#12\catcode `\^12\catcode `\_12\catcode `\%12\relax}%
\providecommand \@@startlink[1]{}%
\providecommand \@@endlink[0]{}%
\providecommand \url  [0]{\begingroup\@sanitize@url \@url }%
\providecommand \@url [1]{\endgroup\@href {#1}{\urlprefix }}%
\providecommand \urlprefix  [0]{URL }%
\providecommand \Eprint [0]{\href }%
\providecommand \doibase [0]{http://dx.doi.org/}%
\providecommand \selectlanguage [0]{\@gobble}%
\providecommand \bibinfo  [0]{\@secondoftwo}%
\providecommand \bibfield  [0]{\@secondoftwo}%
\providecommand \translation [1]{[#1]}%
\providecommand \BibitemOpen [0]{}%
\providecommand \bibitemStop [0]{}%
\providecommand \bibitemNoStop [0]{.\EOS\space}%
\providecommand \EOS [0]{\spacefactor3000\relax}%
\providecommand \BibitemShut  [1]{\csname bibitem#1\endcsname}%
\let\auto@bib@innerbib\@empty
\bibitem [{\citenamefont {Newman}(2003)}]{newman2003structure}%
  \BibitemOpen
  \bibfield  {author} {\bibinfo {author} {\bibfnamefont {M.~E.}\ \bibnamefont
  {Newman}},\ }\href@noop {} {\bibfield  {journal} {\bibinfo  {journal} {SIAM
  review}\ }\textbf {\bibinfo {volume} {45}},\ \bibinfo {pages} {167} (\bibinfo
  {year} {2003})}\BibitemShut {NoStop}%
\bibitem [{\citenamefont {Chen}\ and\ \citenamefont
  {Guanrong}(2012)}]{Chen2012Introduction}%
  \BibitemOpen
  \bibfield  {author} {\bibinfo {author} {\bibnamefont {Chen}}\ and\ \bibinfo
  {author} {\bibnamefont {Guanrong}},\ }\href@noop {} {\emph {\bibinfo {title}
  {Introduction to complex networks : Models, structures and dynamics}}}\
  (\bibinfo  {publisher} {Higher Education Press},\ \bibinfo {year}
  {2012})\BibitemShut {NoStop}%
\bibitem [{\citenamefont {Girvan}\ and\ \citenamefont
  {Newman}(2002)}]{ISI:000176217700005}%
  \BibitemOpen
  \bibfield  {author} {\bibinfo {author} {\bibfnamefont {M.}~\bibnamefont
  {Girvan}}\ and\ \bibinfo {author} {\bibfnamefont {M.~E.}\ \bibnamefont
  {Newman}},\ }\href {\doibase 10.1073/pnas.122653799} {\bibfield  {journal}
  {\bibinfo  {journal} {Proceedings of The National Academy of Sciences}\
  }\textbf {\bibinfo {volume} {99}},\ \bibinfo {pages} {7821} (\bibinfo {year}
  {2002})}\BibitemShut {NoStop}%
\bibitem [{\citenamefont {Newman}(2006)}]{newman2006modularity}%
  \BibitemOpen
  \bibfield  {author} {\bibinfo {author} {\bibfnamefont {M.~E.}\ \bibnamefont
  {Newman}},\ }\href@noop {} {\bibfield  {journal} {\bibinfo  {journal}
  {Proceedings of the National Academy of Sciences}\ }\textbf {\bibinfo
  {volume} {103}},\ \bibinfo {pages} {8577} (\bibinfo {year}
  {2006})}\BibitemShut {NoStop}%
\bibitem [{\citenamefont {Scott}(2017)}]{scott2017social}%
  \BibitemOpen
  \bibfield  {author} {\bibinfo {author} {\bibfnamefont {J.}~\bibnamefont
  {Scott}},\ }\href@noop {} {\emph {\bibinfo {title} {Social network
  analysis}}}\ (\bibinfo  {publisher} {Sage},\ \bibinfo {year}
  {2017})\BibitemShut {NoStop}%
\bibitem [{\citenamefont {Moody}(2004)}]{moody2004structure}%
  \BibitemOpen
  \bibfield  {author} {\bibinfo {author} {\bibfnamefont {J.}~\bibnamefont
  {Moody}},\ }\href@noop {} {\bibfield  {journal} {\bibinfo  {journal}
  {American sociological review}\ }\textbf {\bibinfo {volume} {69}},\ \bibinfo
  {pages} {213} (\bibinfo {year} {2004})}\BibitemShut {NoStop}%
\bibitem [{\citenamefont {Sporns}\ and\ \citenamefont
  {Betzel}(2016)}]{sporns2016modular}%
  \BibitemOpen
  \bibfield  {author} {\bibinfo {author} {\bibfnamefont {O.}~\bibnamefont
  {Sporns}}\ and\ \bibinfo {author} {\bibfnamefont {R.~F.}\ \bibnamefont
  {Betzel}},\ }\href@noop {} {\bibfield  {journal} {\bibinfo  {journal} {Annual
  review of psychology}\ }\textbf {\bibinfo {volume} {67}},\ \bibinfo {pages}
  {613} (\bibinfo {year} {2016})}\BibitemShut {NoStop}%
\bibitem [{\citenamefont {Fortunato}(2010)}]{fortunato2010community}%
  \BibitemOpen
  \bibfield  {author} {\bibinfo {author} {\bibfnamefont {S.}~\bibnamefont
  {Fortunato}},\ }\href@noop {} {\bibfield  {journal} {\bibinfo  {journal}
  {Physics reports}\ }\textbf {\bibinfo {volume} {486}},\ \bibinfo {pages} {75}
  (\bibinfo {year} {2010})}\BibitemShut {NoStop}%
\bibitem [{\citenamefont {Malliaros}\ and\ \citenamefont
  {Vazirgiannis}(2013)}]{malliaros2013clustering}%
  \BibitemOpen
  \bibfield  {author} {\bibinfo {author} {\bibfnamefont {F.~D.}\ \bibnamefont
  {Malliaros}}\ and\ \bibinfo {author} {\bibfnamefont {M.}~\bibnamefont
  {Vazirgiannis}},\ }\href@noop {} {\bibfield  {journal} {\bibinfo  {journal}
  {Physics Reports}\ }\textbf {\bibinfo {volume} {533}},\ \bibinfo {pages} {95}
  (\bibinfo {year} {2013})}\BibitemShut {NoStop}%
\bibitem [{\citenamefont {Newman}(2016)}]{newman2016equivalence}%
  \BibitemOpen
  \bibfield  {author} {\bibinfo {author} {\bibfnamefont {M.~E.}\ \bibnamefont
  {Newman}},\ }\href@noop {} {\bibfield  {journal} {\bibinfo  {journal}
  {Physical Review E}\ }\textbf {\bibinfo {volume} {94}},\ \bibinfo {pages}
  {052315} (\bibinfo {year} {2016})}\BibitemShut {NoStop}%
\bibitem [{\citenamefont {von Luxburg}(2007)}]{ISI:000249409100009}%
  \BibitemOpen
  \bibfield  {author} {\bibinfo {author} {\bibfnamefont {U.}~\bibnamefont {von
  Luxburg}},\ }\href {\doibase 10.1007/s11222-007-9033-z} {\bibfield  {journal}
  {\bibinfo  {journal} {Statistics And Computing}\ }\textbf {\bibinfo {volume}
  {17}},\ \bibinfo {pages} {395} (\bibinfo {year} {2007})}\BibitemShut
  {NoStop}%
\bibitem [{\citenamefont {Holme}\ and\ \citenamefont
  {Saram{\"a}ki}(2012)}]{holme2012temporal}%
  \BibitemOpen
  \bibfield  {author} {\bibinfo {author} {\bibfnamefont {P.}~\bibnamefont
  {Holme}}\ and\ \bibinfo {author} {\bibfnamefont {J.}~\bibnamefont
  {Saram{\"a}ki}},\ }\href@noop {} {\bibfield  {journal} {\bibinfo  {journal}
  {Physics Reports}\ }\textbf {\bibinfo {volume} {519}},\ \bibinfo {pages} {97}
  (\bibinfo {year} {2012})}\BibitemShut {NoStop}%
\bibitem [{\citenamefont {Hutchison}\ \emph {et~al.}(2013)\citenamefont
  {Hutchison}, \citenamefont {Womelsdorf}, \citenamefont {Allen}, \citenamefont
  {Bandettini}, \citenamefont {Calhoun}, \citenamefont {Corbetta},
  \citenamefont {Della~Penna}, \citenamefont {Duyn}, \citenamefont {Glover},
  \citenamefont {Gonzalez-Castillo}, \citenamefont {Handwerker}, \citenamefont
  {Keilholz}, \citenamefont {Kiviniemi}, \citenamefont {Leopold}, \citenamefont
  {de~Pasquale}, \citenamefont {Sporns}, \citenamefont {Walter},\ and\
  \citenamefont {Chang}}]{ISI:000322416000030}%
  \BibitemOpen
  \bibfield  {author} {\bibinfo {author} {\bibfnamefont {R.~M.}\ \bibnamefont
  {Hutchison}}, \bibinfo {author} {\bibfnamefont {T.}~\bibnamefont
  {Womelsdorf}}, \bibinfo {author} {\bibfnamefont {E.~A.}\ \bibnamefont
  {Allen}}, \bibinfo {author} {\bibfnamefont {P.~A.}\ \bibnamefont
  {Bandettini}}, \bibinfo {author} {\bibfnamefont {V.~D.}\ \bibnamefont
  {Calhoun}}, \bibinfo {author} {\bibfnamefont {M.}~\bibnamefont {Corbetta}},
  \bibinfo {author} {\bibfnamefont {S.}~\bibnamefont {Della~Penna}}, \bibinfo
  {author} {\bibfnamefont {J.~H.}\ \bibnamefont {Duyn}}, \bibinfo {author}
  {\bibfnamefont {G.~H.}\ \bibnamefont {Glover}}, \bibinfo {author}
  {\bibfnamefont {J.}~\bibnamefont {Gonzalez-Castillo}}, \bibinfo {author}
  {\bibfnamefont {D.~A.}\ \bibnamefont {Handwerker}}, \bibinfo {author}
  {\bibfnamefont {S.}~\bibnamefont {Keilholz}}, \bibinfo {author}
  {\bibfnamefont {V.}~\bibnamefont {Kiviniemi}}, \bibinfo {author}
  {\bibfnamefont {D.~A.}\ \bibnamefont {Leopold}}, \bibinfo {author}
  {\bibfnamefont {F.}~\bibnamefont {de~Pasquale}}, \bibinfo {author}
  {\bibfnamefont {O.}~\bibnamefont {Sporns}}, \bibinfo {author} {\bibfnamefont
  {M.}~\bibnamefont {Walter}}, \ and\ \bibinfo {author} {\bibfnamefont
  {C.}~\bibnamefont {Chang}},\ }\href@noop {} {\bibfield  {journal} {\bibinfo
  {journal} {{NEUROIMAGE}}\ }\textbf {\bibinfo {volume} {{80}}} (\bibinfo
  {year} {{2013}})}\BibitemShut {NoStop}%
\bibitem [{\citenamefont {Pfitzner}\ \emph {et~al.}(2013)\citenamefont
  {Pfitzner}, \citenamefont {Scholtes}, \citenamefont {Garas}, \citenamefont
  {Tessone},\ and\ \citenamefont {Schweitzer}}]{pfitzner2013betweenness}%
  \BibitemOpen
  \bibfield  {author} {\bibinfo {author} {\bibfnamefont {R.}~\bibnamefont
  {Pfitzner}}, \bibinfo {author} {\bibfnamefont {I.}~\bibnamefont {Scholtes}},
  \bibinfo {author} {\bibfnamefont {A.}~\bibnamefont {Garas}}, \bibinfo
  {author} {\bibfnamefont {C.~J.}\ \bibnamefont {Tessone}}, \ and\ \bibinfo
  {author} {\bibfnamefont {F.}~\bibnamefont {Schweitzer}},\ }\href@noop {}
  {\bibfield  {journal} {\bibinfo  {journal} {Physical review letters}\
  }\textbf {\bibinfo {volume} {110}},\ \bibinfo {pages} {198701} (\bibinfo
  {year} {2013})}\BibitemShut {NoStop}%
\bibitem [{\citenamefont {Zhang}\ and\ \citenamefont
  {Li}(2014)}]{zhang2014susceptible}%
  \BibitemOpen
  \bibfield  {author} {\bibinfo {author} {\bibfnamefont {Y.-Q.}\ \bibnamefont
  {Zhang}}\ and\ \bibinfo {author} {\bibfnamefont {X.}~\bibnamefont {Li}},\
  }\href@noop {} {\bibfield  {journal} {\bibinfo  {journal} {EPL (Europhysics
  Letters)}\ }\textbf {\bibinfo {volume} {108}},\ \bibinfo {pages} {28006}
  (\bibinfo {year} {2014})}\BibitemShut {NoStop}%
\bibitem [{\citenamefont {Zhang}\ \emph {et~al.}(2015)\citenamefont {Zhang},
  \citenamefont {Li}, \citenamefont {Xu},\ and\ \citenamefont
  {Vasilakos}}]{zhang2015human}%
  \BibitemOpen
  \bibfield  {author} {\bibinfo {author} {\bibfnamefont {Y.-Q.}\ \bibnamefont
  {Zhang}}, \bibinfo {author} {\bibfnamefont {X.}~\bibnamefont {Li}}, \bibinfo
  {author} {\bibfnamefont {J.}~\bibnamefont {Xu}}, \ and\ \bibinfo {author}
  {\bibfnamefont {A.~V.}\ \bibnamefont {Vasilakos}},\ }\href@noop {} {\bibfield
   {journal} {\bibinfo  {journal} {IEEE Transactions on Systems, Man, and
  Cybernetics: Systems}\ }\textbf {\bibinfo {volume} {45}},\ \bibinfo {pages}
  {214} (\bibinfo {year} {2015})}\BibitemShut {NoStop}%
\bibitem [{\citenamefont {Hou}\ \emph {et~al.}(2016)\citenamefont {Hou},
  \citenamefont {Li},\ and\ \citenamefont {Chen}}]{hou2016structural}%
  \BibitemOpen
  \bibfield  {author} {\bibinfo {author} {\bibfnamefont {B.}~\bibnamefont
  {Hou}}, \bibinfo {author} {\bibfnamefont {X.}~\bibnamefont {Li}}, \ and\
  \bibinfo {author} {\bibfnamefont {G.}~\bibnamefont {Chen}},\ }\href@noop {}
  {\bibfield  {journal} {\bibinfo  {journal} {IEEE Trans. on Circuits and
  Systems}\ }\textbf {\bibinfo {volume} {63}},\ \bibinfo {pages} {1771}
  (\bibinfo {year} {2016})}\BibitemShut {NoStop}%
\bibitem [{\citenamefont {Liang}\ \emph {et~al.}(2016)\citenamefont {Liang},
  \citenamefont {Li},\ and\ \citenamefont {Zhang}}]{liang2016identifying}%
  \BibitemOpen
  \bibfield  {author} {\bibinfo {author} {\bibfnamefont {D.}~\bibnamefont
  {Liang}}, \bibinfo {author} {\bibfnamefont {X.}~\bibnamefont {Li}}, \ and\
  \bibinfo {author} {\bibfnamefont {Y.-Q.}\ \bibnamefont {Zhang}},\ }\href@noop
  {} {\bibfield  {journal} {\bibinfo  {journal} {EPL (Europhysics Letters)}\
  }\textbf {\bibinfo {volume} {116}},\ \bibinfo {pages} {18006} (\bibinfo
  {year} {2016})}\BibitemShut {NoStop}%
\bibitem [{\citenamefont {Li}\ and\ \citenamefont
  {Li}(2017)}]{li2017reconstruction}%
  \BibitemOpen
  \bibfield  {author} {\bibinfo {author} {\bibfnamefont {X.}~\bibnamefont
  {Li}}\ and\ \bibinfo {author} {\bibfnamefont {X.}~\bibnamefont {Li}},\
  }\href@noop {} {\bibfield  {journal} {\bibinfo  {journal} {Nature
  communications}\ }\textbf {\bibinfo {volume} {8}},\ \bibinfo {pages} {15729}
  (\bibinfo {year} {2017})}\BibitemShut {NoStop}%
\bibitem [{\citenamefont {Aslak}\ \emph {et~al.}(2018)\citenamefont {Aslak},
  \citenamefont {Rosvall},\ and\ \citenamefont
  {Lehmann}}]{aslak2018constrained}%
  \BibitemOpen
  \bibfield  {author} {\bibinfo {author} {\bibfnamefont {U.}~\bibnamefont
  {Aslak}}, \bibinfo {author} {\bibfnamefont {M.}~\bibnamefont {Rosvall}}, \
  and\ \bibinfo {author} {\bibfnamefont {S.}~\bibnamefont {Lehmann}},\
  }\href@noop {} {\bibfield  {journal} {\bibinfo  {journal} {Physical Review
  E}\ }\textbf {\bibinfo {volume} {97}},\ \bibinfo {pages} {062312} (\bibinfo
  {year} {2018})}\BibitemShut {NoStop}%
\bibitem [{\citenamefont {Peixoto}(2015)}]{peixoto2015inferring}%
  \BibitemOpen
  \bibfield  {author} {\bibinfo {author} {\bibfnamefont {T.~P.}\ \bibnamefont
  {Peixoto}},\ }\href@noop {} {\bibfield  {journal} {\bibinfo  {journal}
  {Physical Review E}\ }\textbf {\bibinfo {volume} {92}},\ \bibinfo {pages}
  {042807} (\bibinfo {year} {2015})}\BibitemShut {NoStop}%
\bibitem [{\citenamefont {Ghasemian}\ \emph {et~al.}(2016)\citenamefont
  {Ghasemian}, \citenamefont {Zhang}, \citenamefont {Clauset}, \citenamefont
  {Moore},\ and\ \citenamefont {Peel}}]{ghasemian2016detectability}%
  \BibitemOpen
  \bibfield  {author} {\bibinfo {author} {\bibfnamefont {A.}~\bibnamefont
  {Ghasemian}}, \bibinfo {author} {\bibfnamefont {P.}~\bibnamefont {Zhang}},
  \bibinfo {author} {\bibfnamefont {A.}~\bibnamefont {Clauset}}, \bibinfo
  {author} {\bibfnamefont {C.}~\bibnamefont {Moore}}, \ and\ \bibinfo {author}
  {\bibfnamefont {L.}~\bibnamefont {Peel}},\ }\href@noop {} {\bibfield
  {journal} {\bibinfo  {journal} {Physical Review X}\ }\textbf {\bibinfo
  {volume} {6}},\ \bibinfo {pages} {031005} (\bibinfo {year}
  {2016})}\BibitemShut {NoStop}%
\bibitem [{\citenamefont {Mucha}\ \emph {et~al.}(2010)\citenamefont {Mucha},
  \citenamefont {Richardson}, \citenamefont {Macon}, \citenamefont {Porter},\
  and\ \citenamefont {Onnela}}]{mucha2010community}%
  \BibitemOpen
  \bibfield  {author} {\bibinfo {author} {\bibfnamefont {P.~J.}\ \bibnamefont
  {Mucha}}, \bibinfo {author} {\bibfnamefont {T.}~\bibnamefont {Richardson}},
  \bibinfo {author} {\bibfnamefont {K.}~\bibnamefont {Macon}}, \bibinfo
  {author} {\bibfnamefont {M.~A.}\ \bibnamefont {Porter}}, \ and\ \bibinfo
  {author} {\bibfnamefont {J.-P.}\ \bibnamefont {Onnela}},\ }\href@noop {}
  {\bibfield  {journal} {\bibinfo  {journal} {Science}\ }\textbf {\bibinfo
  {volume} {328}},\ \bibinfo {pages} {876} (\bibinfo {year}
  {2010})}\BibitemShut {NoStop}%
\bibitem [{\citenamefont {Zhang}\ \emph {et~al.}(2012)\citenamefont {Zhang},
  \citenamefont {Zhao},\ and\ \citenamefont {Zhang}}]{zhang2012common}%
  \BibitemOpen
  \bibfield  {author} {\bibinfo {author} {\bibfnamefont {S.}~\bibnamefont
  {Zhang}}, \bibinfo {author} {\bibfnamefont {J.}~\bibnamefont {Zhao}}, \ and\
  \bibinfo {author} {\bibfnamefont {X.-S.}\ \bibnamefont {Zhang}},\ }\href@noop
  {} {\bibfield  {journal} {\bibinfo  {journal} {Physical Review E}\ }\textbf
  {\bibinfo {volume} {85}},\ \bibinfo {pages} {056110} (\bibinfo {year}
  {2012})}\BibitemShut {NoStop}%
\bibitem [{\citenamefont {Zhang}\ and\ \citenamefont
  {Cao}(2017)}]{zhang2017finding}%
  \BibitemOpen
  \bibfield  {author} {\bibinfo {author} {\bibfnamefont {J.}~\bibnamefont
  {Zhang}}\ and\ \bibinfo {author} {\bibfnamefont {J.}~\bibnamefont {Cao}},\
  }\href@noop {} {\bibfield  {journal} {\bibinfo  {journal} {Journal of the
  American Statistical Association}\ }\textbf {\bibinfo {volume} {112}},\
  \bibinfo {pages} {994} (\bibinfo {year} {2017})}\BibitemShut {NoStop}%
\bibitem [{\citenamefont {Bassett}\ \emph {et~al.}(2014)\citenamefont
  {Bassett}, \citenamefont {Wymbs}, \citenamefont {Porter}, \citenamefont
  {Mucha},\ and\ \citenamefont {Grafton}}]{bassett2014cross}%
  \BibitemOpen
  \bibfield  {author} {\bibinfo {author} {\bibfnamefont {D.~S.}\ \bibnamefont
  {Bassett}}, \bibinfo {author} {\bibfnamefont {N.~F.}\ \bibnamefont {Wymbs}},
  \bibinfo {author} {\bibfnamefont {M.~A.}\ \bibnamefont {Porter}}, \bibinfo
  {author} {\bibfnamefont {P.~J.}\ \bibnamefont {Mucha}}, \ and\ \bibinfo
  {author} {\bibfnamefont {S.~T.}\ \bibnamefont {Grafton}},\ }\href@noop {}
  {\bibfield  {journal} {\bibinfo  {journal} {Chaos: An Interdisciplinary
  Journal of Nonlinear Science}\ }\textbf {\bibinfo {volume} {24}},\ \bibinfo
  {pages} {013112} (\bibinfo {year} {2014})}\BibitemShut {NoStop}%
\bibitem [{\citenamefont {Figlewski}(1997)}]{figlewski1997forecasting}%
  \BibitemOpen
  \bibfield  {author} {\bibinfo {author} {\bibfnamefont {S.}~\bibnamefont
  {Figlewski}},\ }\href@noop {} {\bibfield  {journal} {\bibinfo  {journal}
  {Financial markets, institutions \& instruments}\ }\textbf {\bibinfo {volume}
  {6}},\ \bibinfo {pages} {1} (\bibinfo {year} {1997})}\BibitemShut {NoStop}%
\bibitem [{\citenamefont {Newman}\ and\ \citenamefont
  {Girvan}(2004)}]{Newman2004}%
  \BibitemOpen
  \bibfield  {author} {\bibinfo {author} {\bibfnamefont {M.~E.}\ \bibnamefont
  {Newman}}\ and\ \bibinfo {author} {\bibfnamefont {M.}~\bibnamefont
  {Girvan}},\ }\href@noop {} {\bibfield  {journal} {\bibinfo  {journal}
  {Physical review E}\ }\textbf {\bibinfo {volume} {69}},\ \bibinfo {pages}
  {026113} (\bibinfo {year} {2004})}\BibitemShut {NoStop}%
\bibitem [{\citenamefont {Kumpula}\ \emph {et~al.}(2007)\citenamefont
  {Kumpula}, \citenamefont {Onnela}, \citenamefont {Saram{\"a}ki},
  \citenamefont {Kaski},\ and\ \citenamefont
  {Kert{\'e}sz}}]{kumpula2007emergence}%
  \BibitemOpen
  \bibfield  {author} {\bibinfo {author} {\bibfnamefont {J.~M.}\ \bibnamefont
  {Kumpula}}, \bibinfo {author} {\bibfnamefont {J.-P.}\ \bibnamefont {Onnela}},
  \bibinfo {author} {\bibfnamefont {J.}~\bibnamefont {Saram{\"a}ki}}, \bibinfo
  {author} {\bibfnamefont {K.}~\bibnamefont {Kaski}}, \ and\ \bibinfo {author}
  {\bibfnamefont {J.}~\bibnamefont {Kert{\'e}sz}},\ }\href@noop {} {\bibfield
  {journal} {\bibinfo  {journal} {Physical Review Letters}\ }\textbf {\bibinfo
  {volume} {99}},\ \bibinfo {pages} {228701} (\bibinfo {year}
  {2007})}\BibitemShut {NoStop}%
\bibitem [{\citenamefont {Molloy}\ and\ \citenamefont
  {Reed}(1995)}]{molloy1995critical}%
  \BibitemOpen
  \bibfield  {author} {\bibinfo {author} {\bibfnamefont {M.}~\bibnamefont
  {Molloy}}\ and\ \bibinfo {author} {\bibfnamefont {B.}~\bibnamefont {Reed}},\
  }\href@noop {} {\bibfield  {journal} {\bibinfo  {journal} {Random structures
  \& algorithms}\ }\textbf {\bibinfo {volume} {6}},\ \bibinfo {pages} {161}
  (\bibinfo {year} {1995})}\BibitemShut {NoStop}%
\bibitem [{\citenamefont {Newman}\ \emph {et~al.}(2001)\citenamefont {Newman},
  \citenamefont {Strogatz},\ and\ \citenamefont {Watts}}]{newman2001random}%
  \BibitemOpen
  \bibfield  {author} {\bibinfo {author} {\bibfnamefont {M.~E.}\ \bibnamefont
  {Newman}}, \bibinfo {author} {\bibfnamefont {S.~H.}\ \bibnamefont
  {Strogatz}}, \ and\ \bibinfo {author} {\bibfnamefont {D.~J.}\ \bibnamefont
  {Watts}},\ }\href@noop {} {\bibfield  {journal} {\bibinfo  {journal}
  {Physical review E}\ }\textbf {\bibinfo {volume} {64}},\ \bibinfo {pages}
  {026118} (\bibinfo {year} {2001})}\BibitemShut {NoStop}%
\bibitem [{\citenamefont {Blondel}\ \emph {et~al.}(2008)\citenamefont
  {Blondel}, \citenamefont {Guillaume}, \citenamefont {Lambiotte},\ and\
  \citenamefont {Lefebvre}}]{blondel2008fast}%
  \BibitemOpen
  \bibfield  {author} {\bibinfo {author} {\bibfnamefont {V.~D.}\ \bibnamefont
  {Blondel}}, \bibinfo {author} {\bibfnamefont {J.-L.}\ \bibnamefont
  {Guillaume}}, \bibinfo {author} {\bibfnamefont {R.}~\bibnamefont
  {Lambiotte}}, \ and\ \bibinfo {author} {\bibfnamefont {E.}~\bibnamefont
  {Lefebvre}},\ }\href@noop {} {\bibfield  {journal} {\bibinfo  {journal}
  {Journal of Statistical Mechanics: Theory and Experiment}\ }\textbf {\bibinfo
  {volume} {2008}},\ \bibinfo {pages} {P10008} (\bibinfo {year}
  {2008})}\BibitemShut {NoStop}%
\bibitem [{\citenamefont {Bassett}\ \emph {et~al.}(2011)\citenamefont
  {Bassett}, \citenamefont {Wymbs}, \citenamefont {Porter}, \citenamefont
  {Mucha}, \citenamefont {Carlson},\ and\ \citenamefont
  {Grafton}}]{bassett2011dynamic}%
  \BibitemOpen
  \bibfield  {author} {\bibinfo {author} {\bibfnamefont {D.~S.}\ \bibnamefont
  {Bassett}}, \bibinfo {author} {\bibfnamefont {N.~F.}\ \bibnamefont {Wymbs}},
  \bibinfo {author} {\bibfnamefont {M.~A.}\ \bibnamefont {Porter}}, \bibinfo
  {author} {\bibfnamefont {P.~J.}\ \bibnamefont {Mucha}}, \bibinfo {author}
  {\bibfnamefont {J.~M.}\ \bibnamefont {Carlson}}, \ and\ \bibinfo {author}
  {\bibfnamefont {S.~T.}\ \bibnamefont {Grafton}},\ }\href@noop {} {\bibfield
  {journal} {\bibinfo  {journal} {Proceedings of the National Academy of
  Sciences}\ }\textbf {\bibinfo {volume} {108}},\ \bibinfo {pages} {7641}
  (\bibinfo {year} {2011})}\BibitemShut {NoStop}%
\bibitem [{\citenamefont {Shapiro}\ and\ \citenamefont
  {Wilk}(1965)}]{shapiro1965analysis}%
  \BibitemOpen
  \bibfield  {author} {\bibinfo {author} {\bibfnamefont {S.~S.}\ \bibnamefont
  {Shapiro}}\ and\ \bibinfo {author} {\bibfnamefont {M.~B.}\ \bibnamefont
  {Wilk}},\ }\href@noop {} {\bibfield  {journal} {\bibinfo  {journal}
  {Biometrika}\ }\textbf {\bibinfo {volume} {52}},\ \bibinfo {pages} {591}
  (\bibinfo {year} {1965})}\BibitemShut {NoStop}%
\bibitem [{\citenamefont {Mann}\ and\ \citenamefont
  {Whitney}(1947)}]{mann1947test}%
  \BibitemOpen
  \bibfield  {author} {\bibinfo {author} {\bibfnamefont {H.~B.}\ \bibnamefont
  {Mann}}\ and\ \bibinfo {author} {\bibfnamefont {D.~R.}\ \bibnamefont
  {Whitney}},\ }\href@noop {} {\bibfield  {journal} {\bibinfo  {journal} {The
  Annals of Mathematical Statistics}\ ,\ \bibinfo {pages} {50}} (\bibinfo
  {year} {1947})}\BibitemShut {NoStop}%
\bibitem [{\citenamefont {Danon}\ \emph {et~al.}(2005)\citenamefont {Danon},
  \citenamefont {Diaz-Guilera}, \citenamefont {Duch},\ and\ \citenamefont
  {Arenas}}]{danon2005comparing}%
  \BibitemOpen
  \bibfield  {author} {\bibinfo {author} {\bibfnamefont {L.}~\bibnamefont
  {Danon}}, \bibinfo {author} {\bibfnamefont {A.}~\bibnamefont {Diaz-Guilera}},
  \bibinfo {author} {\bibfnamefont {J.}~\bibnamefont {Duch}}, \ and\ \bibinfo
  {author} {\bibfnamefont {A.}~\bibnamefont {Arenas}},\ }\href@noop {}
  {\bibfield  {journal} {\bibinfo  {journal} {Journal of Statistical Mechanics:
  Theory and Experiment}\ }\textbf {\bibinfo {volume} {2005}},\ \bibinfo
  {pages} {P09008} (\bibinfo {year} {2005})}\BibitemShut {NoStop}%
\bibitem [{TP-()}]{TP-toolbox-web}%
  \BibitemOpen
  \href@noop {} {}\bibinfo {howpublished}
  {\url{https://www.senate.gov/legislative/LIS/roll_call_lists/vote_menu_114_1.htm}}\BibitemShut
  {NoStop}%
\bibitem [{Rfm()}]{Rfmriweb}%
  \BibitemOpen
  \href@noop {} {}\bibinfo {howpublished}
  {\url{http://mrirc.psych.ac.cn/DownloadRfMRIMaps}}\BibitemShut {NoStop}%
\bibitem [{adh()}]{adhdweb}%
  \BibitemOpen
  \href@noop {} {}\bibinfo {howpublished}
  {\url{http://fcon_1000.projects.nitrc.org/indi/adhd200/}}\BibitemShut
  {NoStop}%
\bibitem [{\citenamefont {Cheng}\ \emph {et~al.}(2015)\citenamefont {Cheng},
  \citenamefont {Palaniyappan}, \citenamefont {Li}, \citenamefont {Kendrick},
  \citenamefont {Zhang}, \citenamefont {Luo}, \citenamefont {Liu},
  \citenamefont {Yu}, \citenamefont {Deng}, \citenamefont {Wang} \emph
  {et~al.}}]{cheng2015voxel}%
  \BibitemOpen
  \bibfield  {author} {\bibinfo {author} {\bibfnamefont {W.}~\bibnamefont
  {Cheng}}, \bibinfo {author} {\bibfnamefont {L.}~\bibnamefont {Palaniyappan}},
  \bibinfo {author} {\bibfnamefont {M.}~\bibnamefont {Li}}, \bibinfo {author}
  {\bibfnamefont {K.~M.}\ \bibnamefont {Kendrick}}, \bibinfo {author}
  {\bibfnamefont {J.}~\bibnamefont {Zhang}}, \bibinfo {author} {\bibfnamefont
  {Q.}~\bibnamefont {Luo}}, \bibinfo {author} {\bibfnamefont {Z.}~\bibnamefont
  {Liu}}, \bibinfo {author} {\bibfnamefont {R.}~\bibnamefont {Yu}}, \bibinfo
  {author} {\bibfnamefont {W.}~\bibnamefont {Deng}}, \bibinfo {author}
  {\bibfnamefont {Q.}~\bibnamefont {Wang}},  \emph {et~al.},\ }\href@noop {}
  {\bibfield  {journal} {\bibinfo  {journal} {NPJ Schizophrenia}\ }\textbf
  {\bibinfo {volume} {1}},\ \bibinfo {pages} {15016} (\bibinfo {year}
  {2015})}\BibitemShut {NoStop}%
\bibitem [{\citenamefont {Tzourio-Mazoyer}\ \emph {et~al.}(2002)\citenamefont
  {Tzourio-Mazoyer}, \citenamefont {Landeau}, \citenamefont {Papathanassiou},
  \citenamefont {Crivello}, \citenamefont {Etard}, \citenamefont {Delcroix},
  \citenamefont {Mazoyer},\ and\ \citenamefont
  {Joliot}}]{tzourio2002automated}%
  \BibitemOpen
  \bibfield  {author} {\bibinfo {author} {\bibfnamefont {N.}~\bibnamefont
  {Tzourio-Mazoyer}}, \bibinfo {author} {\bibfnamefont {B.}~\bibnamefont
  {Landeau}}, \bibinfo {author} {\bibfnamefont {D.}~\bibnamefont
  {Papathanassiou}}, \bibinfo {author} {\bibfnamefont {F.}~\bibnamefont
  {Crivello}}, \bibinfo {author} {\bibfnamefont {O.}~\bibnamefont {Etard}},
  \bibinfo {author} {\bibfnamefont {N.}~\bibnamefont {Delcroix}}, \bibinfo
  {author} {\bibfnamefont {B.}~\bibnamefont {Mazoyer}}, \ and\ \bibinfo
  {author} {\bibfnamefont {M.}~\bibnamefont {Joliot}},\ }\href@noop {}
  {\bibfield  {journal} {\bibinfo  {journal} {Neuroimage}\ }\textbf {\bibinfo
  {volume} {15}},\ \bibinfo {pages} {273} (\bibinfo {year} {2002})}\BibitemShut
  {NoStop}%
\bibitem [{\citenamefont {Leonardi}\ and\ \citenamefont {Van
  De~Ville}(2015)}]{leonardi2015spurious}%
  \BibitemOpen
  \bibfield  {author} {\bibinfo {author} {\bibfnamefont {N.}~\bibnamefont
  {Leonardi}}\ and\ \bibinfo {author} {\bibfnamefont {D.}~\bibnamefont {Van
  De~Ville}},\ }\href@noop {} {\bibfield  {journal} {\bibinfo  {journal}
  {Neuroimage}\ }\textbf {\bibinfo {volume} {104}},\ \bibinfo {pages} {430}
  (\bibinfo {year} {2015})}\BibitemShut {NoStop}%
\bibitem [{\citenamefont {Nu{\~n}ez-Garcia}\ \emph {et~al.}(2015)\citenamefont
  {Nu{\~n}ez-Garcia}, \citenamefont {Simpraga}, \citenamefont {Jurado},
  \citenamefont {Garolera}, \citenamefont {Pueyo},\ and\ \citenamefont
  {Igual}}]{nunez2015fadr}%
  \BibitemOpen
  \bibfield  {author} {\bibinfo {author} {\bibfnamefont {M.}~\bibnamefont
  {Nu{\~n}ez-Garcia}}, \bibinfo {author} {\bibfnamefont {S.}~\bibnamefont
  {Simpraga}}, \bibinfo {author} {\bibfnamefont {M.~A.}\ \bibnamefont
  {Jurado}}, \bibinfo {author} {\bibfnamefont {M.}~\bibnamefont {Garolera}},
  \bibinfo {author} {\bibfnamefont {R.}~\bibnamefont {Pueyo}}, \ and\ \bibinfo
  {author} {\bibfnamefont {L.}~\bibnamefont {Igual}},\ }in\ \href@noop {}
  {\emph {\bibinfo {booktitle} {International Workshop on Machine Learning in
  Medical Imaging}}}\ (\bibinfo {organization} {Springer},\ \bibinfo {year}
  {2015})\ pp.\ \bibinfo {pages} {61--68}\BibitemShut {NoStop}%
\end{thebibliography}%
\end{document}